\pdfoutput=1

\documentclass[hidelinks,12pt]{article}

\usepackage{hyperref, url} 
\usepackage{graphicx,amsfonts,psfrag,layout,subcaption,array,longtable,lscape,booktabs,dcolumn,amsmath,amssymb,amssymb,amsthm,setspace,epigraph,chronology,color,colortbl,wasysym,diagbox,colortbl,authblk,commath,bbm,natbib}
\usepackage[]{graphicx}\usepackage[]{color}
\usepackage[page]{appendix}
\usepackage[section]{placeins}
\usepackage[linewidth=1pt]{mdframed}

\usepackage{threeparttable}

\usepackage{titlesec} 
\titlelabel{\thetitle.\quad}

\newcommand{\blind}{1}

\if1\blind
{
\title{State-Building through Public Land Disposal? An Application of Matrix Completion for Counterfactual Prediction} 
\author[ ]{Jason Poulos\thanks{\emph{Corresponding Author.} \url{jvpoulos@bwh.harvard.edu}. Division of Endocrinology, Brigham and Women's Hospital, 221 Longwood Avenue, Boston, MA 02115.}}
\affil[ ]{Dept. of Health Care Policy}
\affil[ ]{Harvard Medical School}
\date{\today}
\setcounter{Maxaffil}{0}

}
\fi

\if0\blind
{
	\title{State-Building through Public Land Disposal? An Application of Matrix Completion for Counterfactual Prediction} 
		\author[ ]{}
		\date{}
} \fi

\usepackage{xr}
\usepackage[bottom]{footmisc}

\usepackage{mathtools,tikz,caption}
\DeclareRobustCommand\sampleline[1]{%
	\tikz\draw[#1] (0,0) (0,\the\dimexpr\fontdimen22\textfont2\relax)
	-- (2em,\the\dimexpr\fontdimen22\textfont2\relax);%
}

\usetikzlibrary{plotmarks}

\usepackage{multicol}

\usepackage{xcolor}

\definecolor{Darjeeling11}{HTML}{FF0000}
\definecolor{Darjeeling15}{HTML}{5BBCD6}
\definecolor{darkblue}{rgb}{0.0, 0.0, 0.55}

\definecolor{DID}{HTML}{F8766D}
\definecolor{MC}{HTML}{A3A500}
\definecolor{MC-W}{HTML}{00BF7D}
\definecolor{SCM}{HTML}{00B0F6}
\definecolor{SCM-L1}{HTML}{E76BF3}

\definecolor{Ryellow}{HTML}{fbd46d}
\definecolor{Rorange}{HTML}{fb7813}
\definecolor{Rblue}{HTML}{1b6ca8}

\definecolor{Rdarkred}{HTML}{ff0000}
\definecolor{Rlightred}{HTML}{ffcccc}

\definecolor{Rred}{HTML}{d32626} 
\definecolor{Rlightblue}{HTML}{ADD8E6}

\usepackage{footmisc}
\DefineFNsymbols{mySymbols}{{\ensuremath\dagger}{\ensuremath\ddagger}\S\P
   *{**}{\ensuremath{\dagger\dagger}}{\ensuremath{\ddagger\ddagger}}}
\setfnsymbol{mySymbols}

\usepackage{tabularx}
\newcolumntype{Y}{>{\raggedleft\arraybackslash}X}

\definecolor{Gray}{gray}{0.9}

\newcommand{\captionfonts}{\normalsize}

\makeatletter  
\long\def\@makecaption#1#2{%
  \vskip\abovecaptionskip
  \sbox\@tempboxa{{\captionfonts #1: #2}}%
  \ifdim \wd\@tempboxa >\hsize
    {\captionfonts #1: #2\par}
  \else
    \hbox to\hsize{\hfil\box\@tempboxa\hfil}%
  \fi
  \vskip\belowcaptionskip}
 
\newtheorem*{assumption*}{\assumptionnumber}
\providecommand{\assumptionnumber}{}


\newcommand{\possessivecite}[1]{\citeauthor{#1}'s (\citeyear{#1})} 
\DeclareMathOperator*{\argmin}{arg\,min}

\DeclareMathOperator\bE{\mathbb E} 

\begin{document} 
 
\begin{singlespacing}
\maketitle  
\end{singlespacing}

\thispagestyle{empty}

\begin{abstract}  
\noindent
This paper examines how homestead policies, which opened vast frontier lands for settlement, influenced the development of American frontier states. It uses a treatment propensity-weighted matrix completion model to estimate the counterfactual size of these states without homesteading.  In simulation studies, the method shows lower bias and variance than other estimators, particularly in higher complexity scenarios. The empirical analysis reveals that homestead policies significantly and persistently reduced state government expenditure and revenue. These findings align with continuous difference-in-differences estimates using 1.46 million land patent records. This study's extension of the matrix completion method to include propensity score weighting for causal effect estimation in panel data, especially in staggered treatment contexts, enhances policy evaluation by improving the precision of long-term policy impact assessments.
\end{abstract}	

\pagebreak
\pagenumbering{arabic}

\section{Introduction}

The exploration of state development patterns over time and across regions is a growing area of interest for social scientists. A key contribution in this field comes from \citet[p.~164]{bensel1990}, who emphasizes the significant role of mid-nineteenth century homestead policies --- federal laws aimed at transferring public land to private individuals --- in shaping the developmental trajectory of the United States. Additionally, \citet[p.~250]{murtazashvili2013political} and \citet[p.~12]{frymer2017building} suggest that these policies not only  facilitated land distribution but also enhanced the federal government's bureaucratic capacity to manage public lands and secure future revenue streams. This paper examines how homestead policies impacted the size of state governments, which is closely related to state capacity, or the ability of governments to finance and implement policies \citep{besley2010state}. 

Homesteading is expected to expand the size of state governments by increasing the land values and tax revenue of sparsely populated frontier states. The expansion in state size was historically evident in the adoption of compulsory primary education laws and public education investments by frontier state governments, as a strategy to attract homesteaders \citep{engerman2005evolution}. Contrary to this expectation, I provide evidence that homesteads authorized under the Homestead Act (HSA) of 1862 and the Southern Homestead Act (SHA) of 1866, which opened for settlement hundreds of millions of acres of land for homesteading, significantly reduced the size of frontier state governments over the long-run. The finding that the homestead acts limited the size of frontier governments aligns with \possessivecite{mattheis2019there} findings that regions impacted by the HSA experienced a slower transition from agriculture to other economic sectors and lower housing values. The paper further investigates land inequality as a possible causal mechanism underlying the relationship between homestead policies and state capacity, considering that median voter-based theories of inequality and redistribution predict inequality increases the size of governments, and show that exposure to homesteads decreased land inequality over time. 

The paper makes a methodological contribution in extending the matrix completion method \citep{athey2021matrix} for estimating the causal effects of policy interventions in panel data, by weighting the loss function with estimated unit- and time-varying treatment probabilities (i.e., the propensity score) to correct for imbalances in the covariate distributions between the factual and counterfactual values. This extension, which was proposed by \citet{athey2021matrix} but has not been implemented, places more emphasis on the loss for factual unit-time values that are most similar to the counterfactual values in terms of pre-treatment covariates. The covariates used in the application control for selective migration to more agriculturally productive land, and for selection bias arising from differences in access to frontier lands.

A standard method for causal inference with panel data is difference-in-differences (DID), which relies on the parallel trends assumption: in the absence of treatment, the average outcomes of treated and control units would have followed parallel paths. Under parallel trends, DID identifies causal effects by contrasting the change in outcomes pre- and post-treatment, between the treated and control groups. However, the parallel trends assumption is generally invalid in the presence of unobserved time-varying confounders. DID has been extended to staggered treatment implementation settings, where the time of initial treatment varies among multiple treated units \citep{Callaway2020,goodman2021difference,athey2021design}.

Another popular method of handling unobserved time-varying confounders in panel data is the synthetic control method (SCM; \citealp{abadie2010synthetic}). The method constructs a convex combination of control units that are similar to a single treated unit in terms of pre-treatment outcomes or covariates, to help balance unobserved time-varying confounding between treatment and control groups. The
SCM estimator assumes there is a stable convex combination of the control
units that absorbs all time-varying unobserved confounding. The convexity restriction is
equivalent to imposing a restriction of linear dependence between factor loadings in the
context of matrix completion or latent factor models \citep{gobillon2016regional,xu2017generalized,xiong2020large,bai2021matrix}. The SCM can be generalized to settings with multiple treated units \citep{doudchenko2016balancing} and staggered treatments \citep{benmichael2019synthetic}, and to include features of DID estimation \citep{ben2018augmented,arkhangelsky2021synthetic} or matrix factorization \citep{amjad2018robust,agarwal2021robustness,fan2021exploit}.

Similar to latent factor models, matrix completion attempts to model unobserved
time-varying confounders by decomposing the factual outcomes into matrices of latent factors (i.e., time-varying coefficients) and factor loadings (i.e., unit-specific intercepts). The counterfactual values are then imputed using the estimated factors and loadings. Matrix completion and latent factor models avoid imposing convexity constraints on the factor loadings like the SCM, and typically use matrix norm regularization or factorization to produce a low-dimensional representation of the factual outcomes, which improves generalizability. Matrix completion offers distinct advantages over latent factor models: first, it does not require fixing the rank (i.e., number of unobserved factors) of the underlying data; second, it is suitable in staggered treatment implementation settings, even when few control units are available; third, it uses all factual data to estimate unobserved factors, while latent factor models use only the pre-treatment data.

The structure of this paper is organized as follows: Section \ref{history} provides an overview the historical context of homestead policies and their relationship to state size and land inequality. Section \ref{mc-estimation} details the matrix completion method for obtaining the causal estimands of interest and reports the results of simulation studies to evaluate the proposed method. Section \ref{state-capacity} describes the data sources used for the application and potential sources of bias in the analysis. Section \ref{main-estimates} presents estimates of the long-run impacts of homestead policies on state size using the matrix completion estimator and reports the results of a ``no-treatment evaluation'' to verify the consistency of the estimator. Section~\ref{DID} reports DID estimates of the effect of homesteads on state size and land inequality. The final section discusses the study's findings, focusing on the long-term negative effects of homestead policies on state finances and the role of land inequality in state capacity. It also emphasizes the study's methodological advancement in policy evaluation through the matrix completion method with propensity score weighting.

\setcounter{section}{1} 
\section{Historical background} \label{history}

The view that the western frontier had long-lasting impacts on the evolution of democratic institutions can be traced to \citet{turner1956significance}. Turner's frontier thesis posits that homestead policies acted as a ``safety valve'' for relieving pressure from congested urban labor markets in eastern states. The view of the frontier as a safety valve has been further explored by \citet{ferrie1997migration}, who finds evidence in a linked census sample of substantial migration to the frontier by unskilled workers and considerable gains in wealth for these migrant workers. \citet{bazzi2020frontier} expand on this demographic profile, showing that frontier settlers, often illiterate and foreign-born, possessed a distinct individualism suited for the challenging frontier life. The historical experience of the frontier is reflected in modern times through lower property tax rates in counties with a longer frontier history and a prevailing sentiment among residents against taxation and redistribution.

Homestead policies not only offered greater economic opportunities to eastern migrants, but also the sparse population on the western frontier meant that state and local governments competed with each other to attract migrants in order to lower local labor costs and to increase land values and tax revenue. Frontier governments offered migrants broad access to cheap land and property rights, unrestricted voting rights, and more generous provision of schooling and other public goods \citep{engerman2005evolution}. Consistent with this view, \citet{poulos2021rnn} estimate that the long-run impact of homestead policies on public school spending is equivalent to 2.5\% of the total per-capita  public school expenditures in 1929.

\citet{garcia2008myth} test the frontier thesis in a global context and conclude that the economic effect of the frontier depends on the quality of political institutions at the time of frontier expansion. Frontier expansion promotes equitable outcomes only when societies are initially democratic. When institutional quality is weak, the existence of frontier land can yield worse developmental outcomes because non-democratic political elites can monopolize frontier lands.

\subsection{Homestead policies}\label{homestead-policies}

The 1862 HSA opened up hundreds of millions of acres of western public land for settlement. Any adult household head --- including women, immigrants who had applied for citizenship, and freed slaves following the passage of the Fourteenth Amendment---  could apply for a homestead grant of 160 acres of land, provided that they live and make improvements on the land for five years and pay a \$10 filing fee. Under the HSA, the bulk of newly surveyed land on the western frontier was reserved for homesteads, although the law did not end sales of public land. The explicit goal of the HSA was to liberalize the homesteading requirements set by the Preemption Act of 1841, which permitted individuals already inhabiting public land to purchase up to 160 acres at \$1.25 per acre before the land was put up for sale. The implicit goal was to promote rapid settlement of the western frontier \citep{allen1991homesteading}, and to reduce the federal government's costs of defending contested frontiers \citep{frymer2014rush}.

In the pre-Reconstruction South, public land was not open to homestead but rather unrestricted cash entry, which permitted the direct sale of public land to private individuals of 80 acres or more for at least \$1.25 an acre. The 1866 SHA restricted cash entry and reserved for homesteading over 46 million acres of public land, or about one-third of the total land area in the five southern public land states \citep[p.~13]{lanza1999agrarianism}. The Bureau of Land Management (BLM) classified land disposed under the SHA under the same authority as land disposed by the HSA, since the SHA amended the HSA and dictated that public lands be disposed under the stipulations of the HSA \citep{hoffnagle1970southern}. 

\subsubsection{Public and state-land states}

\begin{figure}[htbp]
	\centering
	\begin{subfigure}[b]{\textwidth}
		\centering
		\includegraphics[width=\textwidth]{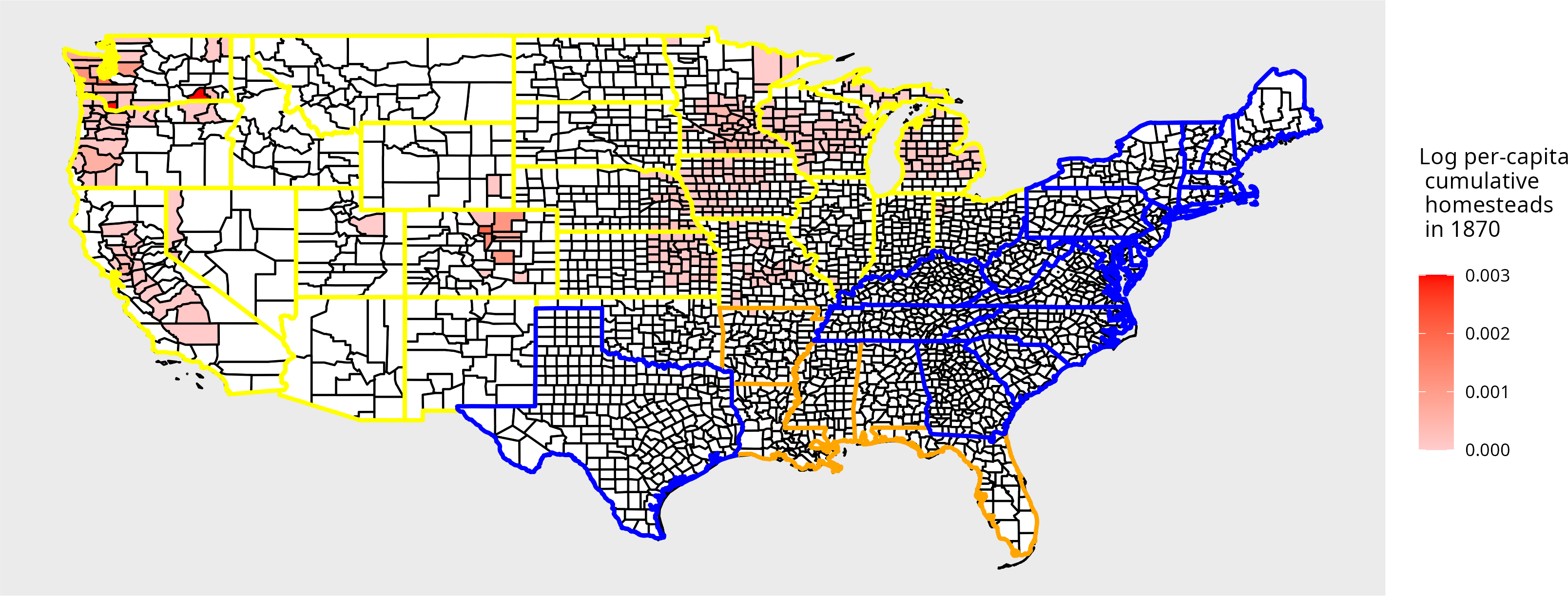}
		\caption{1870}
		\label{fig:1870}
	\end{subfigure}
	\begin{subfigure}[b]{\textwidth}
		\centering
		\includegraphics[width=\textwidth]{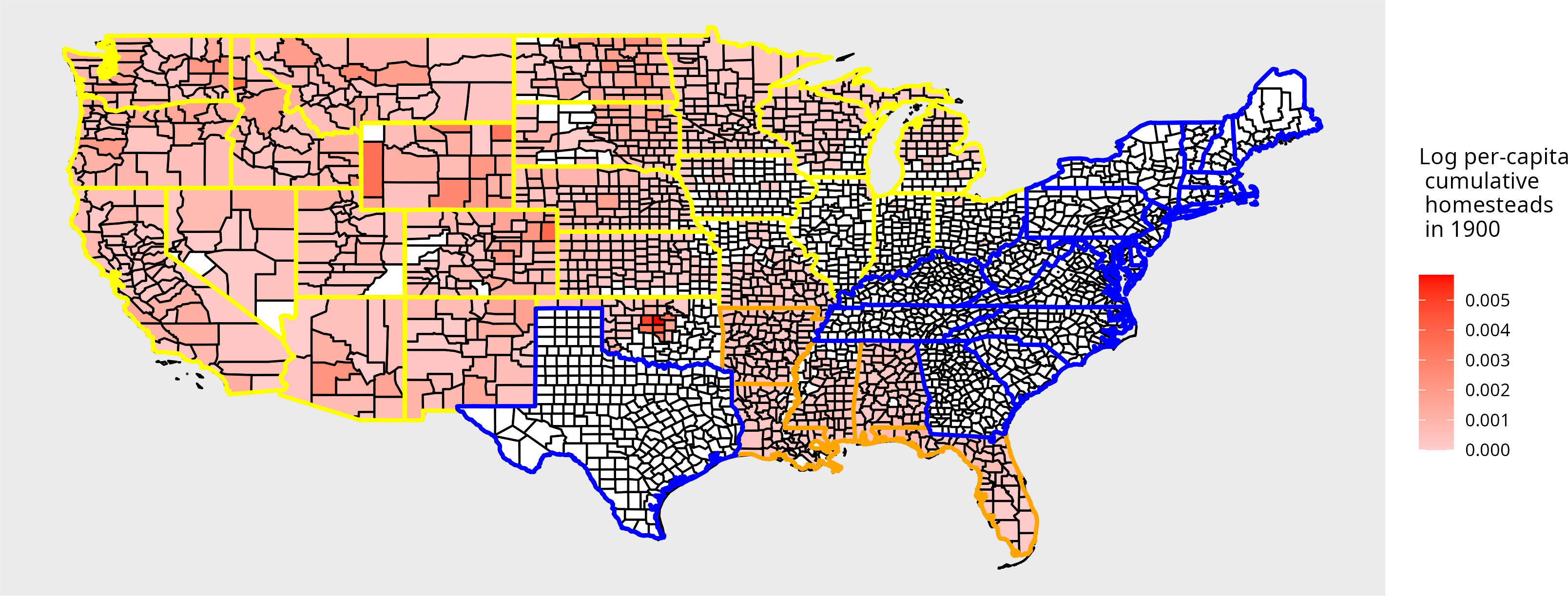}
		\caption{1900}
		\label{fig:1900}
	\end{subfigure}
	\caption{Log per-capita cumulative homesteads in 1870 and 1900, overlaid on 1911 county borders \citep{long1995atlas}. White-colored counties have no homestead entries. States bordered in blue ({\protect\tikz \protect\draw[color={Rblue}] (0,0) -- plot[mark=square, thick, mark options={scale=2}] (0,0);}) are state land states; yellow ({\protect\tikz \protect\draw[color={Ryellow}] (0,0) -- plot[mark=square, thick, mark options={scale=2}] (0,0);})  denote western public land states; orange ({\protect\tikz \protect\draw[color={Rorange}] (0,0) -- plot[mark=square, thick, mark options={scale=2}] (0,0);}) denote southern public land states. \label{homestead-map}} 
\end{figure}

Public land states (PLS) are states that were crafted from the public domain, and where the federal government has the primary authority to distribute public land \citep[p.~4]{murtazashvili2013political}. In the South, these states include Alabama, Arkansas, Florida, Louisiana, and Mississippi. Western PLS include the 25 states that comprise the Midwestern, Southwestern, and Western U.S. (except Hawaii). State-land states, which include the original 13 states, Kentucky, Maine, Tennessee, Texas, Vermont, and West Virginia, were not open to homesteading because the state governments had primary authority to distribute public land. The maps in Figure \ref{homestead-map} reflect the impact of these policies, indicating a significant increase in log per-capita cumulative homesteads in Southern and Western PLS by 1900.

\subsubsection{Challenges and speculation in homesteading}
There were substantial barriers to entry to homesteading, and homesteaders took on enormous risk in the five years required to file a homestead patent. One of the most significant obstacles to entry was the need for capital to build a successful farming operation: contemporary writers estimated that potential homesteaders required \$600 to \$1000 to start a farm \citep{deverell1988loosen}. The high cost associated with starting and maintaining a farm casts doubt on the safety valve hypothesis \citep{danhof1941farm}, and the effectiveness of the land policies such as the HSA as a wealth-building tool was limited by the binding capital constraints faced by small farmers \citep[p.~35]{gates1996jeffersonian}. \citet{poulos2023gender} further contributes to this discussion by examining the 1901 Oklahoma Land Lottery, demonstrating gender differences in leveraging lottery wealth for land purchases and homestead patents, with female winners more effectively using lottery wealth to overcome liquidity constraints in entrepreneurial activities.

Homesteading was a risky venture: over the period of 1910 to 1919, out of 604,092 homestead entries in the U.S., totaling over 128 million acres, only 384,954 (63.7\%) resulted in successful patents \citep{shanks2005homestead}. At least part of the discrepancy between homestead entries and filings, however, may be explained by fraudulent filings. Speculators and corporations engaged in the practice of paying individual to stake a claim in a homestead, with no intention of completing the patent, in order to extract resources from the land \citep{gates1936homestead}. In the South, these ``dummy entrymen'' were used by timber and mining companies to extract resources while the cash entry restriction of the SHA was in effect. When the restriction was removed, there was no need for fraudulent filings because the larger companies could buy land in unlimited amounts at a nominal price \citep{gates1940federal, gates1979federal}. The same pattern of fraudulent filings existed in the West, where \citet[p.~216]{murtazashvili2013political} argues that speculators benefited disproportionately from public land policies because the economic balance of power tilted toward the wealthy. \citet{gates1942role} characterizes western speculators who bought land in bulk prior to the 1889 restriction as being influential in state and local governments, resistant to paying taxes, and opposed to government spending. 

\subsection{Land inequality as a causal mechanism}  \label{mechanisms}

Inequality is a potential causal mechanism underlying the relationship between homesteads and state size. Median voter-based theories that assume parity in the political influence of voters predict a positive relationship between inequality and the size of governments \citep{meltzer1981rational}. In settings with high inequality, the median voter is poorer than the average voter, which in turn increases demand for redistribution in majority-rule elections. 

However, models that allow for economic differences in the political influence of voters predict a nonlinear or inverse relationship between inequality and government size. In \possessivecite{benabou2000unequal} model, for instance, the pivotal voter is wealthier than the median and has the power to block redistribution as inequality increases. But when inequality is too high, the poor can impose redistribution on elites through majority voting \citep{perotti1993political,saint1993education}. In \possessivecite{besley2009origins} framework, for example, greater economic power of the ruling class reduces government spending and investments in state capacity. Similarly, \citet{galor2009inequality} propose a model where wealthy landowners block education reforms because education favors industrial labor productivity and decreases the value in farm rents. Inequality in this context can be thought of as a proxy for the amount of \emph{de facto} political influence elites have to block reforms and limit the size of states \citep{acemoglu2008persistence}. 

To test whether homesteads affected future land inequality in frontier counties, I calculate a commonly-used measure of land inequality based on the Gini coefficient of census farm sizes, adjusted by the ratio of farms to adult males, a measure proposed by \citet{vollrath2013inequality}. Gini-based land inequality measures are commonly used as proxy for the \emph{de facto} bargaining power of landed elites \citep[e.g.,][]{boix2003democracy,ziblatt2008does,ansell2015}. 

Figure \ref{fig:ineq-capacity} in the online Appendix plots the results from bivariate regression models of land inequality and state government finances during the period of 1860 to 1950, demonstrating a positive relationship among groups of PLS and state-land states. This associational evidence is consistent with the predictions of the \citeauthor{meltzer1981rational} model; however, it contrasts with recent empirical studies that establish a negative relationship based on cross-sectional analyses. \citet{ramcharan2010inequality}, for instance, finds an inverse relationship between land inequality and county-level property tax revenue in 1890. The authors find that the negative relationship is especially large in rural counties, where landownership tends to be more concentrated. \citet{vollrath2013inequality} establish a negative relationship between land inequality and local property tax revenue in northern rural counties in 1890. The present findings, in contrast, are based on state-level expenditure and revenue panel data. 

\section{Matrix completion for counterfactual prediction} \label{mc-estimation}

This paper applies the matrix completion method proposed by \citet{athey2021matrix} to predict counterfactual outcomes, and extends the method by propensity-weighting the loss function to correct for imbalances in the factual covariate distributions between treatment and control groups.

\subsection{Setup and notation}\label{notation}
We consider a sample of $i \in \left\{1, \ldots, N\right\}$ units, each observed in $t \in \left\{1, \ldots, T\right\}$ time periods. Following the notation of \citet{athey2021design}, let $\mathbf{a}$ be a length-$N$ vector, where $a_i \in \left\{1, \ldots, T, \infty \right\}$ indexes the time of initial treatment, and $a_i=\infty$ denotes control units. If a unit enters treatment during the panel ($a_i \neq \infty$), it remains treated for the remainder of the panel. There is a nonzero number of control units, $N_{\text{C}} = N - \sum_{i} \mathbbm{1}_{a_{i}=\infty}$, with $\mathbbm{1}(\cdot)$ denoting the indicator function, and a nonzero number treated units $N_{\text{T}} = N- N_{\text{C}} = \sum_{i} \mathbbm{1}_{a_{i} \neq \infty}$. Let the values of the treatment indicator $W_{it} \in \left\{0,1\right\}$ be $W_{it} = 0$ for the control units in all time periods and $W_{it} = 1$ for the treated units when $t \geq a_i$. Let $\mathcal{O}$ denote the set of factual outcome values; i.e., the values for which $W_{it} =0$. 

Under the Neyman-Rubin potential outcomes framework \citep{neyman1923,rubin1990}, for each unit $i$ and time $t$, there exist potential outcomes $Y(a_i)_{it}$. The fundamental problem is that we can only observe a single potential outcome for each unit-time observation: $Y(a_i)_{it}$ is observed for treated units when entering treatment, and $Y(\infty)_{it}$ is observed for the control units in all time periods. The potential outcomes framework implicitly assumes treatment is well-defined to ensure that each unit has the same number of potential outcomes. It also requires that the potential outcomes of unit $i$ varies with $a_i$ but not with the other values of $\mathbf{a}$, which is often referred to as the no interference assumption.

There are two additional assumptions are needed to write potential outcomes as a function of $\mathbf{a}$, which are both made in \citet{athey2021design}. First, there are no anticipatory effects; i.e., $Y(a_i)_{it} = Y(\infty)_{it}$ for all $a_i > t$. This assumption, which is often implicitly made in panel data studies, assumes that if a unit has not yet entered treatment, the initial treatment time has no causal effect on potential outcomes in the current period. Second, potential outcomes in period $t$ are invariant to how long unit $i$ has been exposed to treatment; i.e., $Y(a_i)_{it} = Y(1)_{it}$ for all $a_i \leq t$. This assumption does not rule out causal effects of treatment duration on the outcome, but rather rules out causal effects varying by initial treatment time.

\subsection{Causal estimands}

The causal estimand of interest is the average treatment effect on the treated units (ATT) of entering treatment in $a_{i}^{\prime}$ relative to being control ($a_{i} = \infty$), on the outcome in period $t$:
\begin{equation} 
\tau_{t,  \infty a_{i}^{\prime}} = \frac{1}{N_{T}} \sum_{i=1}^{N_{T}} Y(a_{i}^{\prime})_{it}-Y(\infty)_{it}, \qquad \text{for } W_{it}=1. \label{eq:ATT}
\end{equation}

In the application, I consider $a_i^{\prime} = \min_{1\leq i \leq N_T} a_i$, or the year of the earliest homestead entry among the treated units. The ATT averaged over the counterfactual period, which provides a summary measure of the overall treatment effect, is also of interest:
\begin{equation} 
\tau_{\infty a_{i}^{\prime}} = \frac{1}{T-a_{i}^{\prime}+1}\sum_{t=a_{i}^{\prime}}^{T}\tau_{t,  \infty a_{i}^{\prime}}. \label{eq:ATT-avg}
\end{equation}

\subsection{Estimation}\label{estimation}

In the application, the outcome of interest is state size, measured by state government spending and revenue. The goal is to estimate the potential outcomes under control for the treated units; i.e., the counterfactual state size of treated units had they not been exposed to treatment. I model the outcome under control as:
\begin{equation}
	Y(\infty)_{it} = L_{it} + \gamma_{i} + \delta_{t} + \epsilon_{it}, \label{eq:mc-Y}
\end{equation} 
where $L_{it}$ is a typical element in the unknown matrix, $\mathbf{L} = \mathbf{U} \mathbf{V}^{\top}$, the product of a matrix of factor loadings, $\mathbf{U}_{N \times R}$, and a matrix of factors, $\mathbf{V}_{T \times R}$. While latent factor models assume the rank, or number of unobserved factors $R$, is fixed, matrix completion methods assume that the rank of $\mathbf{L}$ is low relative to $N$ and $T$. The model includes unit-specific fixed effects, $\{\gamma_i\}^N_{i=1}$, and time-specific fixed effects, $\{\delta_t\}^T_{t=1}$, which are meant to capture unobserved confounders not absorbed by the low-rank matrix. The identifying assumption is that the errors $\epsilon_{it}$ are conditionally mean zero and independent of $a_{i}$, for all values of $i$ and $t$:
\begin{equation}\label{exogeneity}
\bE(\epsilon_{it} | L_{it}, \gamma_{i}, \delta_{t}) = \bE(\epsilon_{it} | L_{it}, \gamma_{i}, \delta_{t}, a_{i}) = 0.
\end{equation}
\noindent
This assumption rules out correlation between the errors and initial treatment time in any period \citep{benmichael2019synthetic}. It is analogous to the strict exogeneity assumption made for the estimation of the ATT using latent factor models \citep{xu2017generalized}.

Estimating $\mathbf{L}$ involves minimizing the sum of squared errors via nuclear norm regularized least squares:
\begin{equation}
	\argmin_{\mathbf{L}, \boldsymbol{\gamma}, \boldsymbol{\delta}} \Bigg[\frac{1}{|\mathcal{O}|} \sum_{(i,t) \in \mathcal{O}}   \frac{\widehat{\text{w}}_{it}}{1-\widehat{\text{w}}_{it}} \, \bigg(Y(\infty)_{it} - L_{it} - \gamma_{i} - \delta_{t} \bigg)^2
	+ \lambda_L \norm{\mathbf{L}}_\star \Bigg], \label{eq:mc-opt-prop}
\end{equation}
where the nuclear norm, $\norm{\cdot}_\star = \sum_{i} \sigma_i (\cdot)$, or sum of singular values, is used to yield a low-rank solution for $\mathbf{L}$. The value of the hyperparameter $\lambda_L$ is chosen among 30 possible values by five-fold cross-validation, where in each fold, 80\% of the entries in $\mathcal{O}$ are randomly selected to be used for training, while the remaining 20\% of entries are used for model validation. The model with $\lambda_L$ values that yield the lowest root mean squared error averaged over the validation sets is then fit using all entries in $\mathcal{O}$. 

To quantify the propensity score, $w_{it}$, I model the probability of treatment as:
\begin{equation}\label{treatment-model}
w_{it} =  \mbox{Pr} \left(W_{it} = 1 | Y_{i,1}, \ldots, Y_{i,a_{i}^{\prime}-1} + X_{ip} \right), \qquad  0 < w_{it} < 1,
\end{equation}
where $X_{ip}$ is a typical element in a matrix of $p$ covariates measured prior to $a_{i}^{\prime}$. In the application, the covariates include state-level measures of racial composition, prevalence of pre-emancipation slavery, average farm sizes and values, and railroad access. I estimate the treatment model by multivariate lasso logistic regression \citep{friedman2010regularization}. 

The squared loss in Eq. \eqref{eq:mc-opt-prop} is weighted by estimated propensity scores, $\widehat{w}_{it}$, to place more emphasis on the loss for the values in $\mathcal{O}$ most similar to the counterfactual values in terms of pre-treatment outcomes and covariates. Consistent estimation of the ATT \eqref{eq:ATT} relies on the correct specification of the outcome model \eqref{eq:mc-Y} under the assumption of exogeneity \eqref{exogeneity}. It does not rely on the correct specification of the treatment model \eqref{treatment-model}; although, the propensity scores from estimating this model are intended to help balance treated and control units in terms of the pre-treatment outcomes and covariates when fitting the matrix completion model on the factual data.

The algorithm for solving Eq. \eqref{eq:mc-opt-prop} iteratively replaces missing values with those recovered from a singular value decomposition of the matrix \citep{mazumder2010spectral}. Once $\mathbf{L}$, $\boldsymbol{\gamma}$, and $\boldsymbol{\delta}$ have been estimated, we can predict the counterfactual values for the treated units in the post-treatment period by
\[
\widehat{Y}(\infty)_{it} = \widehat{L}_{it} + \widehat{\gamma}_{i} + \widehat{\delta}_{t}, \qquad \forall \, (i,t) \, \notin \mathcal{O}.
\]
\subsection{Simulation studies}\label{simulation-studies}

In simulation studies described in Section \ref{sims} in the online Appendix, I conduct two sets of simulation studies to assess the performance of the matrix completion estimator: the first on generated data in which we control the ground-truth treatment effects; and the second on empirical data, where the focus is on time periods where no treatment effects are expected. The comparison estimators are evaluated on their ability to recover the ground-truth ATT averaged over the counterfactual period, $\tau_{\infty a_{i}^{\prime}}$ \eqref{eq:ATT-avg}. The comparison estimators include two versions of the matrix completion estimator: with (MC-W) and without (MC) a treatment propensity-weighted loss function. The DID estimator is a regression of outcomes on treatment and unit and time fixed effects. The SCM is a regression of the pre-treatment outcomes of each treated unit on the control unit outcomes during the same periods, with the restrictions of no intercept and non-negative regression weights that sum to one. The SCM with lasso (SCM-L1) relaxes the zero-intercept and weight restrictions and estimates the counterfactual outcomes for each treated unit by lasso regression. I provide the exact form of these estimators in Section \ref{benchmark-estimators}.

\subsubsection{Generated data} 
\begin{figure}[htbp]
	\centering
	\begin{subfigure}[b]{0.49\textwidth}
		\centering
		\includegraphics[width=\textwidth]{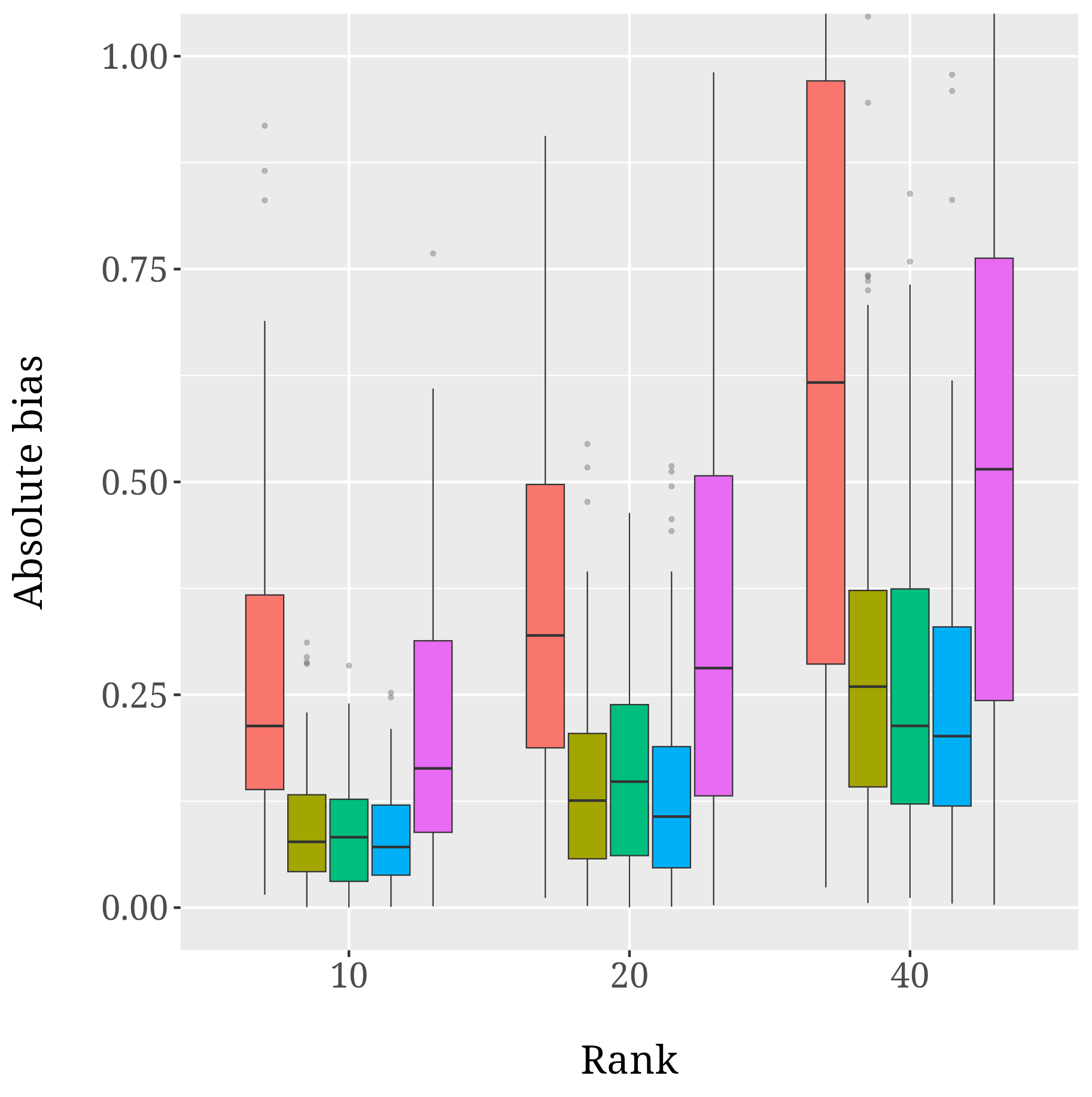}
		\caption{Absolute bias}
		\label{fig:abs_bias}
	\end{subfigure}
	\hfill
	\begin{subfigure}[b]{0.49\textwidth}
		\centering
		\includegraphics[width=\textwidth]{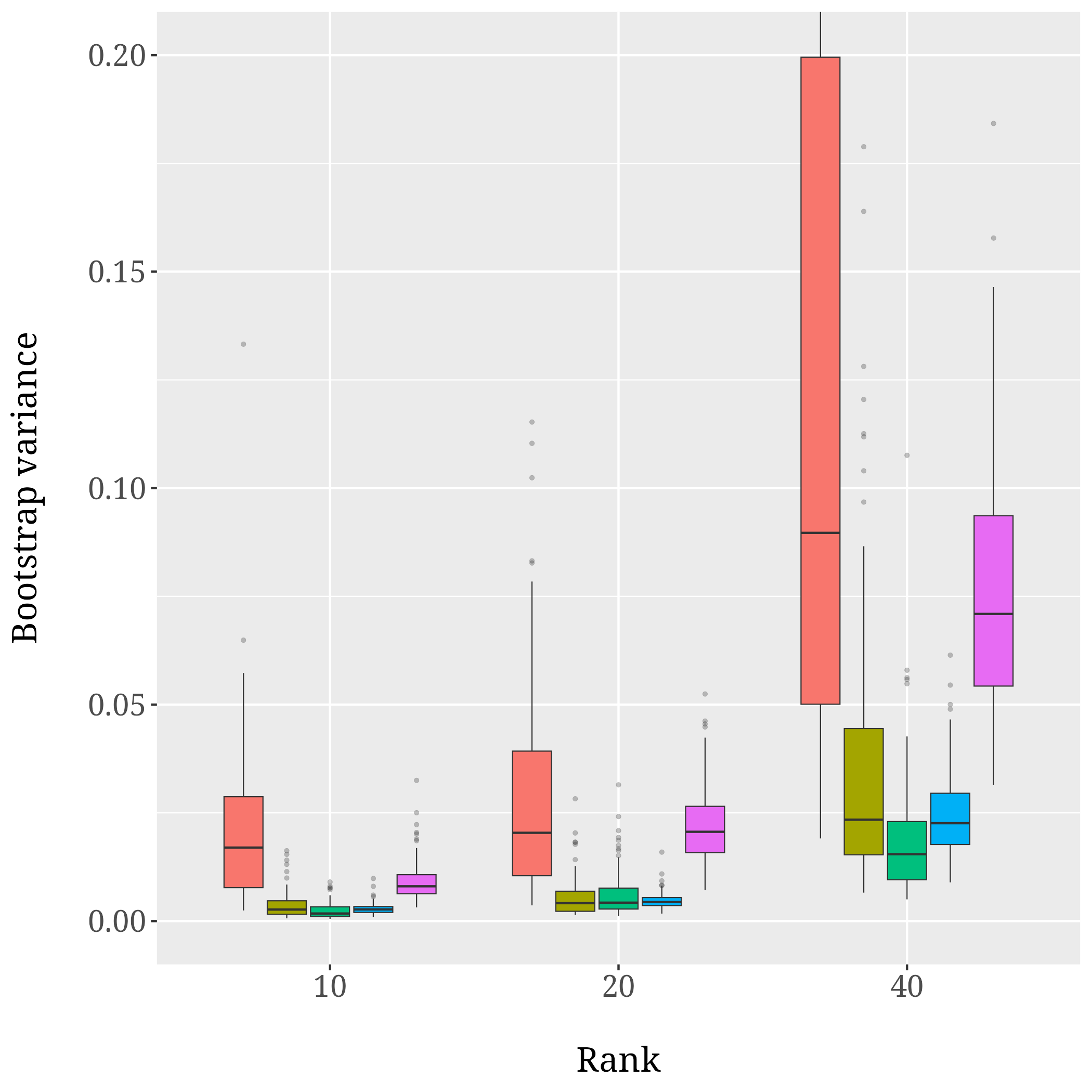}
		\caption{Bootstrap variance}
		\label{fig:boot_var}
	\end{subfigure}
	
	\caption{Absolute bias and bootstrap variance in generated data, varying the rank of $L_{it}$. 
				\emph{Estimator:} 
				{\protect\tikz \protect\draw[color=DID] (0,0) plot[mark=square*, mark options={scale=2.5,rotate=0}] (0.25,0);} DID; 
				{\protect\tikz \protect\draw[color=MC] (0,0) plot[mark=square*, mark options={scale=2.5, rotate=0}] (0.25,0);} MC;
				{\protect\tikz \protect\draw[color=MC-W] (0,0) plot[mark=square*, mark options={scale=2.5, rotate=0}] (0.25,0);} MC-W;
				{\protect\tikz \protect\draw[color=SCM] plot[mark=square*, mark options={scale=2.5, rotate=0}] (0.25,0);} SCM;
				{\protect\tikz \protect\draw[color=SCM-L1] plot[mark=square*, mark options={scale=2.5, rotate=0}] (0.25,0);}, SCM-L1. 
		\label{mc_simulation_placebo}}
\end{figure}

Figure \ref{mc_simulation_placebo} provides box and whisker plots summarizing the median, the first and third quartiles, and outlying points of the distribution of the absolute bias --- i.e., absolute difference between $\widehat{\tau}_{\infty a_{i}^{\prime}}$ and the actual $\tau_{\infty a_{i}^{\prime}}$ --- and the variance of 399 block bootstrap replicates of $\widehat{\tau}_{\infty a_{i}^{\prime}}$ for the first set of simulations on generated data. The absolute bias and bootstrap variance increase across all estimators as the rank of $L_{it}$ increases, which underscores the importance of the low-rank assumption. The matrix completion estimators and the SCM estimator yield the lowest absolute bias and bootstrap variance, regardless of rank, whereas the DID and SCM-L1 estimators struggle with higher bias and variance. When the rank is high relative to $N$ and $T$, MC-W exhibits lower absolute bias and bootstrap variance relative to the unweighted MC estimator, suggesting that propensity score weighting may mitigate the impact of increased model complexity on estimator bias and variance.

\subsubsection{Empirical data} In the second set of simulations focusing on the empirical data, I leverage the fact that the true treatment effect is null in the pre-treatment period. I first discard the post-treatment data, and for each of 1000 simulation runs, randomly select half of the control units to be treated and impute their missing values following a placebo treatment time randomly chosen from $\{a_{i}^{\prime}, \ldots, T\}$, varying $a_{i}^{\prime}$.
	
In simulation studies on the state government finance datasets, described in Section \ref{state-capacity}, the matrix completion estimators generally maintain lower absolute bias than the DID estimator while exhibiting higher bias when compared to SCM estimators, across all placebo $a_{i}^{\prime}$ ratios for both expenditure and revenue datasets (Figure \ref{mc_capacity_simulation_placebo_abs_bias}). In terms of bootstrap variance, matrix completion estimators demonstrate results on par with synthetic control estimators and markedly outperform the DID estimator (Figure \ref{simulation-boot-var}). There is not much efficiency gain from propensity-weighting the matrix completion estimator in the empirical data simulation studies because, unlike the generated data simulations, treatment is assigned at random, rather than as a function of covariates.

In each of three datasets common to the synthetic control literature, the matrix completion estimators outperform DID and the SCM estimators by minimizing absolute bias (Figure \ref{synth_abs_bias}) and bootstrap variance (Figure \ref{synth_boot_variance}) across the different ratios of the placebo $a_{i}^{\prime}$ to $T$. Together, the simulation results support the preferential use of the MC-W estimator in the application. 

\section{Application: Homestead policy and state size} \label{state-capacity}

In order to estimate causal impacts of homestead policies on state size, I create measures of total expenditure and revenue collected from the records of 48 state governments during the period of 1789 to 1932 \citep{sylla1993sources}, 16 state governments during the period of 1933 to 1937 \citep{sylla1995sourcesa,sylla1995sourcesb}, and U.S. Census special reports for the period of 1902 to 2008, covering 48 states \citep{haines2010,census2010}. The expenditure measure includes state government spending on education, social welfare programs, and transportation. The revenue measure incorporates state government income streams such as tax revenue and non-tax revenue such as land sales.

The expenditure and revenue data pre-processing steps are as follows. Removing years with zero or near-zero variance results in outcome matrices consisting of $T=203$ observations for $N=48$ states, 30 of which are treated. The outcome data are inflation-adjusted according to the U.S. Consumer Price Index \citep{williamson2017seven} and scaled by the total free adult male population in the decennial census \citep{haines2010}. I impute the outcome values that are missing due to lack of data collection using multiple imputation by chained equations (MICE, \citealp{buuren2010mice}). Figure \ref{fig:missing-heatmap} visualizes the extent of the missing data in the entire dataset by state and treatment group, where 40\% of values in the dataset are missing (29.9\% and 10.1\% missing in the control and treated groups, respectively). The majority of the outcome data for treated states prior to the treatment time were missing and have been imputed. To address the concern that the choice of imputation method can influence the estimated treatment effects, Table \ref{mc-sens} evaluates the sensitivity of the causal estimates to alternative imputation methods. Lastly, I log-transform the data to alleviate exponential effects.  

The staggered treatment implementation setting is appropriate in this application because $a_i$ varies across states that were exposed to homesteads following the passage of the HSA. I aggregate to the state level approximately 1.46 million individual land patent records authorized under the HSA. Using these records, which are made available by the BLM \citep{GLO}, I determine that the earliest homestead entries occurred in 1869 in about half of the western frontier states, about seven years following the enactment of the HSA. In 1872, the first homesteads were filed in southern PLS. Figure \ref{fig:treat-heatmap} shows how each state is categorized in the empirical analysis, as a PLS (treated group) or state-land state (control group), as well as the year of the earliest initial homestead entry for the PLS, which informs staggered treatment implementation.

I include the following covariates in the conditioning set of the treatment model \eqref{treatment-model}: per-capita spending or revenue prior to 1869; the ratio of slaves to the total population in 1860; and the ratio of free African-Americans, Native Americans, or Whites to the total non-slave population in 1860; average farm sizes in 1860 and average farm values in 1850 and 1860 \citep{haines2010}; and the state-level share of total miles of operational railroad track per square mile, which I calculate by overlaying the railroad track map over historical county borders \citep{atack2013use}. These pre-treatment covariates control for selective migration to more agriculturally productive land, and for differences in the accessibility and availability of frontier lands. \citet[p.~45]{bustos2017essays} finds that the prevalence of slavery in 1860 is an important predictor of available homestead lands, and reasons that the covariate acts as a proxy for the presence of large plantations.

\subsection{Accounting for bias}\label{bias}

Potential sources of bias include violations of the assumptions of exogeneity \eqref{exogeneity}, no interference, no-anticipation, or invariance to treatment history. The exogeniety assumption would be violated if the error term in the outcome model \eqref{eq:mc-Y} is correlated with the initial treatment time. While this assumption is not directly testable, the no-treatment evaluation on pre-treatment data reported in Section \ref{sec-placebo-tests} provide indirect evidence that the exogeneity assumption is not violated. Additionally, the simulation results on the state government finances datasets reported in Section \ref{simulation-studies} demonstrate that propensity-weighting the loss function improves the consistency of the matrix completion estimator.

A second potential source of bias arises from interference, or the assumption that control units are unaffected by the effects of treatment. While the no interference assumption cannot directly be tested, it is likely in the present application that the outcomes of state-land states were indirectly affected by the out-migration of homesteaders from frontier states. When assuming the absence of interference, the use of indirectly affected states as control units would underestimate treatment effects because it would make the counterfactual and factual treated unit observations in the post-treatment period more similar. Interference might also arise if state-land state governments increase public investments in order to dissuade workers from migrating to the frontier in the first place. The historical evidence, however, suggests that labor-scarce frontier states were more strongly motivated to attract migrants and stimulate population growth than long-settled state-land states \citep{engerman2005evolution}. For example, the adoption of compulsory primary education laws and support for public education in general in western states has been considered as a means to attract potential migrants to the frontier \citep{meyer1979public,bandiera2018nation}. Interference arising from competition among state governments would also underestimate the effect of treatment.

A third potential source of bias arises in violations of the no-anticipation or invariance to treatment history assumptions. The no-anticipation assumption would be violated if there were anticipatory effects on the size of frontier state governments prior to the initial homestead entries. Anticipatory effects are plausible since the first homestead entries occurred in 1869 in western PLS, six years after the HSA went into effect. In Section \ref{sec-placebo-tests}, I conduct a no-treatment evaluation on the pre-treatment data and vary the placebo initial treatment year. The estimated placebo ATT is nonsignificant for most settings, which is direct evidence of no anticipatory effects. The invariance to treatment history assumption rules out variation in treatment effects by the initial treatment time, but does not rule out causal effects of treatment duration. In Section \ref{hetero}, I explore whether causal effects on state size differ with respect to year of initial homestead entry.

Lastly, bias may result from misspecification of imputation models. The imputation procedure assumes that after controlling for the available state government finances data, the missing values are Missing At Random (MAR). There are reasons to believe the data are not MAR, which could result in biased estimates. For example, the timing of a state's admission to the Union, which affects the extent of its missing values, may be determined by unobserved political and demographic variables rather than meeting a population threshold. It is impossible to distinguish whether data are MAR or missing based on unobserved variables, given the observed data \citep{sterne2009multiple}. In Section \ref{main-estimates}, the sensitivity of causal estimates to two alternative imputation methods is evaluated, indicating that the choice of imputation method alters the conclusions in one out of the two scenarios examined, as detailed in Table \ref{mc-sens}.

\section{Matrix completion estimates} \label{main-estimates}

\begin{figure}[htbp]
	\centering
	\includegraphics[width=0.9\textwidth]{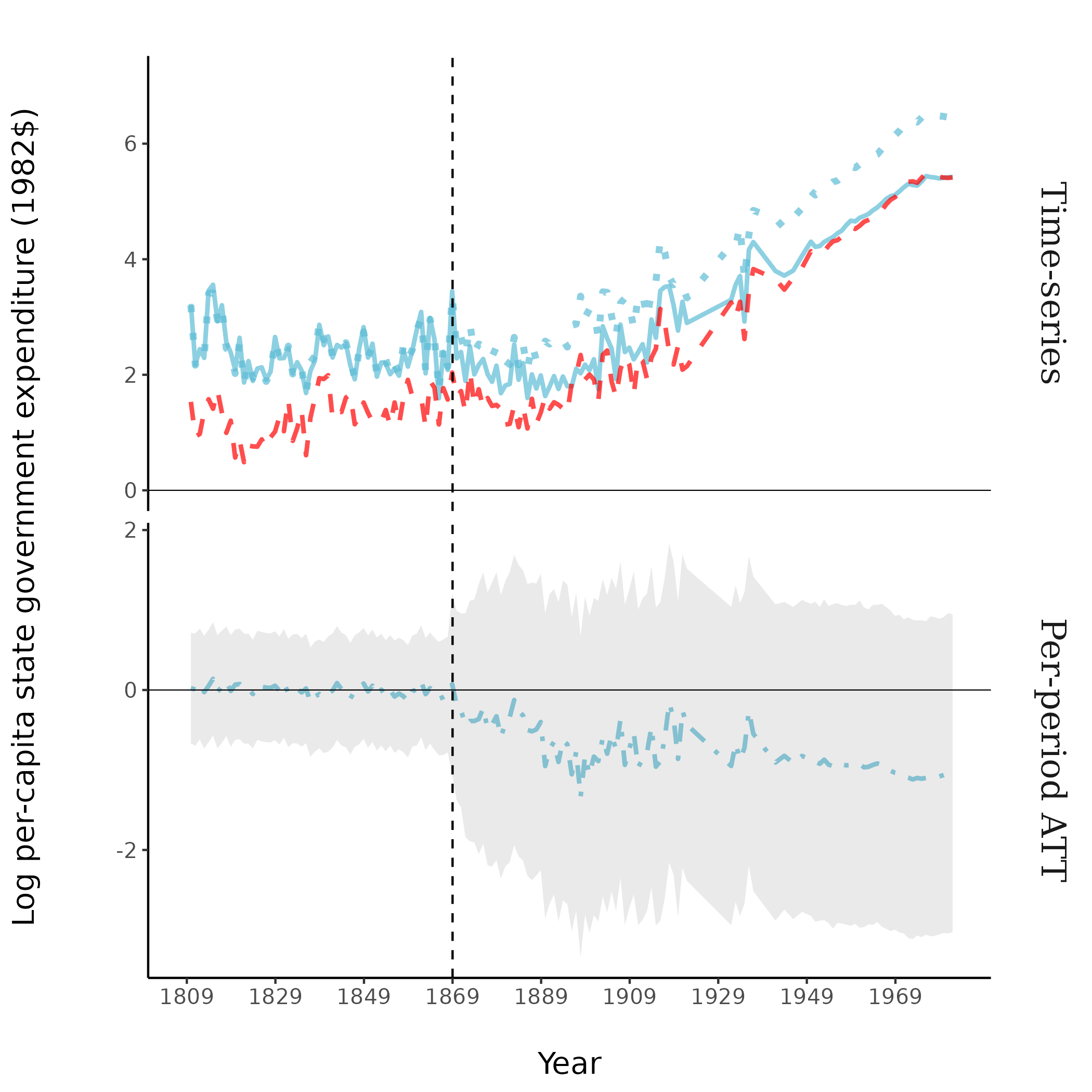}
	\caption{Matrix completion estimates of the effect of the year of initial homestead entry (1869; dashed vertical line) on state government expenditure, 1809 to 1982:
		{\color{Darjeeling15}{\sampleline{line width=1mm}}}, factual treated;
		{\color{Darjeeling11}{\sampleline{dashed,line width=0.75mm}}}, factual control;
		{\color{Darjeeling15}{\sampleline{dotted,line width=0.75mm}}}, counterfactual treated;
		{\color{Darjeeling15}{\sampleline{dash pattern=on .7em off .2em on .05em off .2em,line width=0.75mm}}}, $\widehat{\tau}_{t, \infty a_{i}^{\prime}}$.\label{mc-estimates-exp-pc}} 
\end{figure}

I estimate the causal impacts of the initial treatment year on the state government finances of the treated units (i.e., PLS). Specifically, I fit the MC-W estimator on the entirety of factual outcomes to recover the counterfactual outcomes of the treated units had they not been exposed to treatment. The top panel of Figure \ref{mc-estimates-exp-pc} compares the average log per-capita state government expenditure of treated units and control units along with the predicted average expenditure of treated units. The dashed vertical line represents the year of the earliest homestead entry, $a_{i}^{\prime} = \min_{1\leq i \leq N_T} a_i = 1869$. The treated unit outcomes are generally higher than those of the control units in the pre-treatment period, whereas there is little difference between the treated and control unit outcomes, on average, in the post-treatment period. 

The difference between the observed and predicted treated unit outcomes, which is the quantity $\widehat{\tau}_{t, \infty a_{i}^{\prime}}$ described in Eq. \eqref{eq:ATT}, corresponds to the estimated per-period ATT. These per-period average causal impacts are plotted in the bottom panel, with 95\% normal interval confidence intervals estimated by calculating the standard error of the distribution of block bootstrap replicates of $\widehat{\tau}_{t, \infty a_{i}^{\prime}}$. The bootstrap replicates are constructed by block resampling the columns (i.e, time dimension) of the observed outcomes, in order to preserve temporal dependence structure of the original data \citep{davison1997bootstrap,politis2004automatic}, and obtaining a set of point estimates from 999 resamples. The estimated per-period effects for both outcomes are essentially zero during the pre-treatment period and within the bounds of the bootstrap confidence intervals, which demonstrates that the model is closely fitting the pre-treatment period observations. By 1876, after most PLS had been exposed to homesteads, homestead exposure decreases per-capita state government expenditure by about 0.13 log points, and the trajectory of estimated causal impacts remains negative for the rest of the time-series, although the confidence intervals for the per-period effects all contain zero. Similar patterns are observed when the outcome is per-capita state government revenue (Figure \ref{mc-estimates-rev-pc}).

To infer the overall effect of treatment, I estimate the ATT averaged over the counterfactual period of 1869 to 2008, and report the point estimates and bootstrap standard errors in the second and third columns of Table \ref{mc-estimates}. The estimates reveal that per-capita state government expenditure would have been 0.84 [0.61, 1.08] log points lower had the PLS never been exposed to homesteads. Relative to the observed log per-capita state expenditure of the PLS in the same period, the point estimate represents a decrease of 0.22\%. The estimated ATT on per-capita state government revenue is similar.

I compare the MC-W estimates with DID and SCM estimates, also reported in the second and third columns of Table \ref{mc-estimates}. Figure \ref{did-pretest} illustrates the parallel trends assumption in DID analysis by depicting the log per-capita state government expenditures and revenues over time for both treated and control groups, showing similar trajectories up to the treatment year of 1869, which supports the validity of the DID approach. The point estimates from the binary DID estimator are slightly larger and within the confidence intervals of the MC-W estimates. The ATT estimates from the SCM estimator are positive, but not statistically significant.

Table \ref{mc-sens} presents counterfactual period estimates on differently imputed datasets. When estimated on data with missing outcome values imputed by MICE with classification and regression trees (CART) as the imputation method, rather than predictive mean matching (the default method), the conclusions drawn from the estimates do not change. However, when estimating on data with missing values imputed by an expectation--maximization (EM) algorithm based method, the MC-W estimates are much larger in magnitude and no longer statistically significant. In interpreting the results presented in Table \ref{mc-sens}, it is important to reflect on the implications of imputing a majority of the outcomes in treated states prior to the treatment period. The differences in ATT estimates, especially under the EM imputation method, underscore the sensitivity of our results to the imputation of missing data.

\begin{table}[htbp]
	\captionsetup{font=normalsize}
	\caption{Estimates of the ATT averaged over the counterfactual period of 1869 to 2008 and bootstrap standard errors (in parentheses).\label{mc-estimates}}
	\begin{center}
		\begin{tabular}{@{}lcccccc@{}}
	\toprule 
	& \multicolumn{2}{c}{\emph{All PLS} } & \multicolumn{2}{c}{ \emph{Southern PLS}  } & \multicolumn{2}{c}{ \emph{Western PLS}  }\\
	& \multicolumn{1}{c}{ Expenditure } & \multicolumn{1}{c}{ Revenue } & \multicolumn{1}{c}{ Expenditure } & \multicolumn{1}{c}{ Revenue } & \multicolumn{1}{c}{ Expenditure } & \multicolumn{1}{c}{ Revenue } \\
	DID  & -0.90 (0.32) &  -0.95 (0.27)	& -0.69 (0.39)	& -0.55 (0.35)	&-0.94 (0.45)	& -1.04 (0.51)	\\
	MC-W  & -0.84 (0.24) & -0.79 (0.25) & -0.63 (0.30)	& -0.42 (0.30)	& -0.88 (0.37)	&	-0.86 (0.39)\\
	SCM  & 0.13 (0.12) & 0.17 (0.14) & -0.09 (0.21)	& -0.14 (0.23)	& 0.17 (0.26)	& 0.23 (0.27) \\
	SCM-L1 & 0.13 (0.17)   & 0.17 (0.19) & -0.08 (0.23)	& -0.11 (0.21)	& 0.17 (0.27)	& 0.22 (0.29)	\\
	\bottomrule
\end{tabular}

	\end{center}
\end{table}

\subsection{Treatment effect heterogeneity}\label{hetero} 

Recall that under staggered treatment implementation, the time of initial treatment $a_i$ varies across states that were exposed to homesteads, and that the year of the earliest homestead entry among the treated units $a_i^{\prime}$ is used to calculate the ATT. Also recall that the SHA opened land for homesteading in the South under the same stipulations as the HSA, which opened land for homesteading in the western frontier.  The results above set $a_i^{\prime}$ at 1869, which is the earliest homestead entry that occurred in the western PLS. Among southern PLS, the earliest homestead entry occurred in 1872. Thus, there is a substantive interest in determining whether there is a differential effect of the year of initial homestead entry on state size based on region. Conducting a sub-group analysis by region also allows us to detect potential violations in the assumption of invariance to treatment history, since most of the western PLS are treated for a longer period than the southern PLS.

The last four columns of Table \ref{mc-estimates} decomposes the counterfactual period estimates by calculating the ATT with respect to the region of the PLS. The MC-W estimates show that the main effect on all of the PLS (second and third columns) is driven mainly by the effect on the Western PLS. The estimated effect size on the southern PLS are comparatively smaller in magnitude and significant for the effect on state government expenditure, -0.63 [-0.94, -0.33], but not on revenue, -0.42 [-0.73, -0.12]. These results provide indirect evidence that the assumption of invariance to treatment history is not violated since the conclusions drawn from the main estimates are generally unchanged.

\subsection{No-treatment evaluation}\label{sec-placebo-tests}

To assess whether the estimated effects are attributable to the year of initial homestead entry rather than other policy changes or spurious errors during the same period, I conduct a no-treatment evaluation by discarding the post-treatment period observations from the state government finances data and re-running the analysis on the pre-treatment data, when no treatment effect is expected (i.e., the ATT is zero).
	
Table \ref{placebo-tests} reports placebo ATT estimates and block bootstrap standard errors on each outcome, considering $t = \left\{1, \ldots,  a_{i}^{\prime} - \Delta\right\}$ as the pre-treatment period, with $\Delta \in \left\{1, 10, 25\right\}$. Across all estimators, the standard error decreases with a larger $\Delta$, reflecting the uncertainty of estimating causal effects in shorter (placebo) post-treatment periods. Compared to the binary DID and SCM estimators, MC-W exhibits lower standard errors in all settings and lower bias in three of the six settings. The placebo ATT estimates from the MC-W estimator is significant only when the outcome is state government expenditure and $\Delta$ is 10 or 25. Similar patterns are observed when conducting no-treatment evaluations on differently imputed datasets (Table \ref{placebo-tests-sens}). These placebo results bolster the usage of the MC-W estimator in the application, and provide evidence supporting the plausibility of the exogeneity and no-anticipation assumptions.

\begin{table}[htbp]
	\captionsetup{font=normalsize}
	\caption{Placebo ATT estimates and bootstrap standard errors (in parentheses).\label{placebo-tests}}
	\begin{center}
		\scalebox{.9}{\begin{tabular}{@{}lcccccc@{}} 
	\toprule
	& \multicolumn{3}{c}{ Expenditure } & \multicolumn{3}{c}{ Revenue } \\
	  &  $\Delta=1$ & $\Delta=10$ & $\Delta=25$  &  $\Delta=1$ & $\Delta=10$ & $\Delta=25$  \\
	DID  & -0.61 (0.43) & -0.38 (0.24) & -0.39 (0.22) & -0.63 (0.48)  & -0.70 (0.42) & -0.67 (0.34) \\
	MC-W  & -0.26 (0.32) & -0.44 (0.16)& -0.47 (0.15) & -0.33 (0.38)  & -0.26 (0.21) & -0.28 (0.16)  \\
	SCM  & 0.94 (0.45) &  0.86 (0.24) & 0.76 (0.19) & 0.66 (0.47) & 0.52 (0.28) & 0.64 (0.22) \\
	SCM-L1  & 0.48 (0.41) &0.34 (0.22) & 0.48 (0.16) & 0.61 (0.39) & 0.39 (0.26) & 0.20 (0.21)  \\
	\bottomrule
\end{tabular}
}
	\end{center}
\end{table}

\section{Continuous DID estimation} \label{DID}

The matrix completion approach estimates the impact of a binary exposure to treatment on a continuous outcome. In the application, however, a continuous form of treatment is available in the number of homestead patents. The model below is a continuous version of the DID estimator described in Section \ref{did-binary}, where the first difference comes from variation in the date of initial exposure to homesteads, and the second difference comes from variation in the intensity of homestead entries:

\begin{equation} 
Y_{it} =  \xi_i + \psi_t + \zeta \, W_{it} + \phi \, (W_{it} \cdot H_{it}) + \beta X_{ip}  + \upsilon_{it}. \label{eq:dd} 
\end{equation}
The model includes state and year fixed effects, $\{\xi_i\}^N_{i=1}$ and $\{\psi_t\}^T_{t=1}$, respectively. The covariate $X_{ip}$ controls for average farm sizes and values, and railroad access, when the outcome $Y_{it}$ is log per-capita state government spending or revenue; $X_{ip}$ controls for average farm values when $Y_{it}$ is land inequality. The continuous treatment variable $H_{it}$ measures the log of the per-capita number of patents issued under the HSA in state $i$ and year $t$. The coefficient of interest corresponds to the interaction term, $\phi$, which represents the average causal effect of exposure to homesteads. A least squares estimator for $\phi$ is given by 
\begin{equation}  
\argmin_{\phi,\xi_i,\psi_t, \zeta, \beta}
\sum_{i=1}^N \sum_{t=1}^T 
\left(
Y_{it}-\xi_i-\psi_t -\zeta \, W_{it} - \phi \, (W_{it} \cdot H_{it}) - \beta X_{ip}\right)^2. \label{eq:dd-est}
\end{equation}
%

\subsection{Estimates on state size and land inequality}

Table \ref{dd-estimates} reports DID estimates of the average causal effect of exposure to log per-capita homestead patents, with 95\% confidence intervals constructed using 999 state-stratified bootstrap samples. The estimates indicate that a 1\% increase in log per-capita homesteads decreases log per-capita state government spending or revenue by about 4\%. The point estimates are considerably smaller in magnitude --- albeit, in the same direction --- as the per-period MC-W estimates presented in Section \ref{main-estimates}. The bootstrap confidence intervals around the DID estimates are considerably more narrow than those for the MC-W estimates displayed in the bottom panels of Figures \ref{mc-estimates} and \ref{mc-estimates-rev-pc}, and are potentially overoptimistic due to serial correlation in the DID regression errors \citep{bertrand2004much}. The estimates on state size are insensitive to the method used for imputing expenditure or revenue values that are missing due to nonresponse (Table \ref{dd-estimates-sens}).

\begin{table}[htp]
	\caption{Continuous DID estimates of the effect of (log) per-capita homestead patents on state size or land inequality. \label{dd-estimates}}
	\begin{center}
		\resizebox{.9\width}{!}{	\begin{tabular}{@{}lccc@{}}
		\toprule
										& Expenditure             & Revenue                 & Land inequality \\ \hline \\
		Treatment effect ($\hat{\phi}$) &  -0.04				  &  -0.04 				&   -0.0007        \\
										&  (0.002) 	  	  & (0.002)  &   (0.0003)         \\
									    &                   	  &                    		&          \\
		Adjusted $\text{R}^2$           & 0.538            & 0.540                 &   0.801         \\
		$\text{N}$                      & 8,618                   & 8,618                   &  463          \\
		 \bottomrule
	\end{tabular}}
\end{center}
\end{table}

The third column of Table \ref{dd-estimates} presents DID estimates of the impact of log per-capita homesteads on land inequality at the state-level during the period of 1870 to 1950. Since land inequality is measured every decennial, I aggregate homesteads to the next decennial year; e.g., the number of homesteads measured in 1880 is the total for the years 1871 to 1880. Average farm values are included in the regression as a proxy for agricultural productivity, which might be associated with farm sizes approaching ideal scale and therefore land inequality. I estimate that homesteads significantly decreased land inequality in frontier states: a 1\% increase in log per-capita homesteads lowers state-level land inequality by about $10^{-5}$ points. 

The direction of the point estimate is consistent with the study of \citet[p.~3]{bustos2017essays}, who conducts a county-level DID analysis and shows treatment based on terciles of Homestead Act acres reduced land inequality measured by the Gini coefficient over a similar post-treatment period, although the magnitude of the estimate in the present work is substantially smaller. The comparatively small coefficient implies that homestead policies did not fundamentally alter the long-run distribution of landownership, which may be explained by qualitative evidence that suggests homestead policies were exploited by land speculators and natural resource companies and that the rents from public land were appropriated by the private sector. 

\section{Conclusion} \label{discussion} 

The findings of this paper signify that mid-nineteenth century homestead policies had long-lasting impacts that can potentially explain contemporary differences in state government size. Estimates using matrix completion with a binary treatment and DID with continuous treatment evidence that homestead policies had significant and negative impacts on state government expenditure and revenue that lasted a century following their implementation. This finding is in line with recent work documenting the adverse impact of homestead policies on the economic development of regions exposed to homesteading. 

I explore land inequality as a possible causal mechanism underlying the relationship between homestead policies and state size, which is closely related to state capacity. First, I provide evidence of a positive relationship between land inequality and state government finances and that the slope of correlation increases at higher levels of inequality. Nonlinearities in the relationship between inequality and state capacity can arise in theoretical models that incorporate economic differences in political influence: greater income inequality reduces government spending and investments in state capacity when elites have a monopoly on political power, however when inequality gets too high, the poor can impose redistribution through majority voting. Second, I present continuous DID estimates that reveal per-capita homesteads significantly lowered land inequality in frontier states; although, the magnitude of the effect is negligible. This finding is in line with previous empirical work showing that exposure to homesteads decreased land inequality. The failure to fundamentally alter the long-run distribution of landownership may be explained by qualitative evidence that suggests homestead policies were \emph{de facto} corporate welfarism often exploited by land speculators and corporations to amass land and resources during early capitalist expansion. 

This paper makes a methodological contribution by extending the matrix completion method for causal effect estimation in panel data with staggered treatment adoption to allow for propensity score weighting of the loss function. The matrix completion estimator with propensity score weighting outperforms regression-based estimators such as the synthetic control method and difference-in-differences in simulation studies and a no-treatment evaluation. This methodological contribution holds implications for policy evaluation, offering a more accurate tool for understanding the effects of policies over time and place.

\if1\blind
{
\section*{Funding details}

This work was supported by the National Science Foundation under Grants DGE 1106400 and TG-SES 180010.
}
\fi

\section*{Supplemental online material}

The online Appendix includes simulation results, and describes model specifications and implementation details for each of the comparison estimators used in the simulations. It includes descriptive figures on the extent of missing data in the state government finances datasets, and reports the results of sensitivity analyses on differently imputed datasets. It also includes figures for matrix completion estimates of treatment exposure on state government revenue, a diagnostic plot for the DID parallel trends assumption, and bivariate regression estimates of the relationship between land inequality and state government finances.

\if1\blind
{
\section*{Data and code}
Data and \textsf{R} code to reproduce the results of the paper are available at \url{https://github.com/jvpoulos/homesteads}.
}
\fi

\newpage

\bibliographystyle{asa}
\begin{singlespace}
\bibliography{references}
\end{singlespace}

\itemize
\end{document}


\begin{singlespacing}
	\maketitle  
\end{singlespacing}

\thispagestyle{empty}

\begin{abstract}
	\noindent 
	The online Appendix includes maps describing the treatment status of states (Figures \ref{homestead-map} and \ref{fig:treat-heatmap}); bivariate regression estimates of the relationship between land inequality and state government finances (Figure \ref{fig:ineq-capacity}); a write-up of the design and results of the simulation studies (Section \ref{sims}); and the model specifications and implementation details for each of the comparison estimators (Section \ref{benchmark-estimators}). Lastly, Section \ref{tables-figures} includes tables and figures for the extent of missing data in the state government finances datasets (Figure \ref{fig:missing-heatmap}); matrix completion estimates of treatment exposure on state government revenue (Figure \ref{mc-estimates-rev-pc}); a diagnostic plot for the DID parallel trends assumption (Figure \ref{did-pretest}); and reports the results of sensitivity analyses on differently imputed datasets (Table \ref{mc-sens}). 
\end{abstract}


\setcounter{figure}{0} \renewcommand{\thefigure}{A\arabic{figure}}
\setcounter{table}{0} \renewcommand{\thetable}{A\arabic{table}}
\setcounter{section}{0} \renewcommand{\thesection}{A\arabic{section}}

\pagenumbering{roman}
\pagebreak
\pagenumbering{arabic}

\section{Empirical application: Descriptive statistics}

\begin{figure}[htbp]
	\begin{center}
		\includegraphics[width=\textwidth]{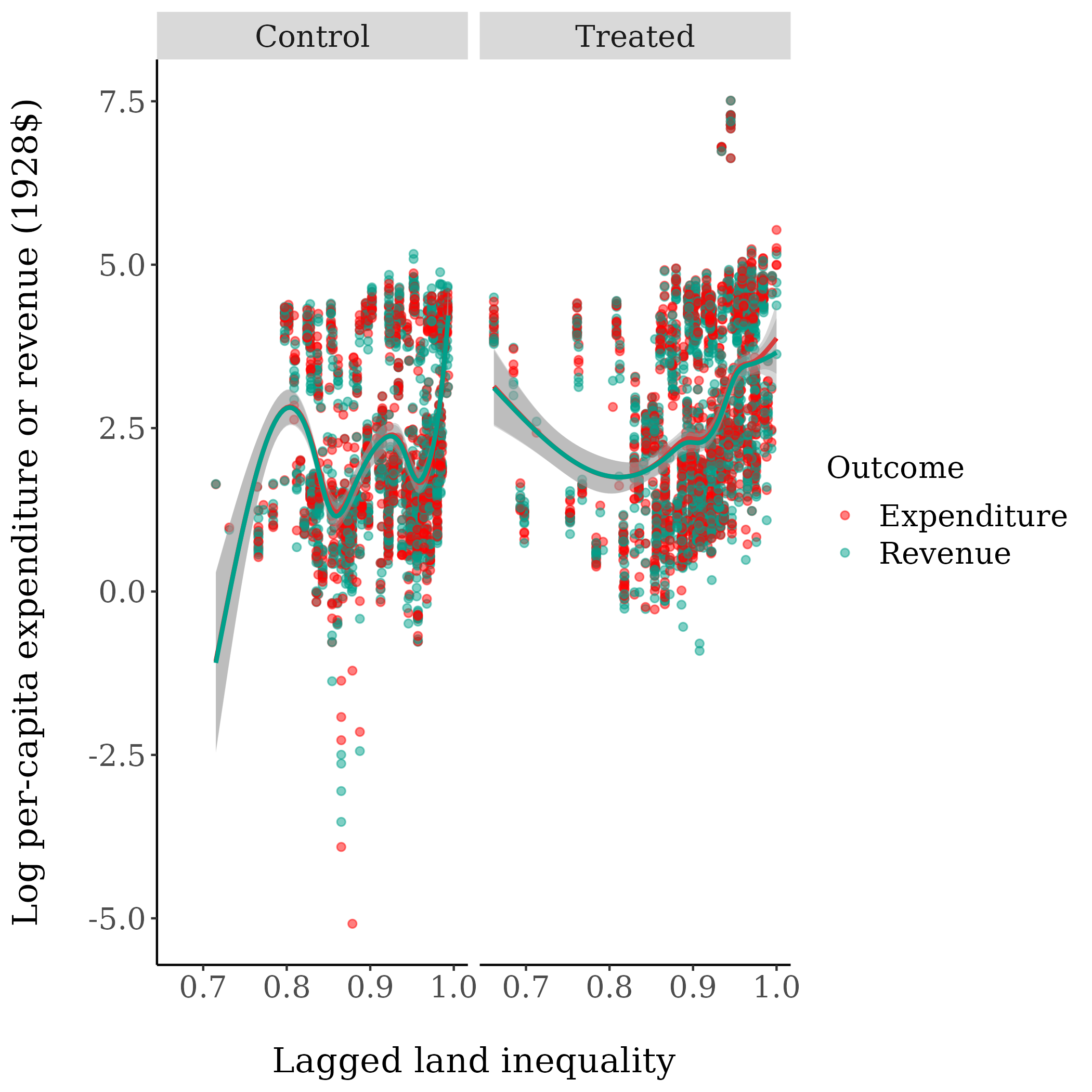} 
	\end{center}
	\caption{Land inequality (lagged by 10 years) vs. log per-capita state government revenue and expenditure, 1860-1950. Each point is a state-year observation. Lines represent generalized additive model fits to the data for the two outcomes and shaded regions represent corresponding 95\% confidence intervals. The model is fit separately on control states (i.e., state-land states) and treated states (i.e., PLS).   \label{fig:ineq-capacity}}
\end{figure}

\section{Simulations} \label{sims}

\subsection{Generated data} \label{sec:simulation-gen}

In the first set of simulations, I generate potential outcomes under control according to the outcome model \eqref{eq:mc-Y}. For each of 100 trial runs, I generate the low-rank matrix $L_{it}$ as the product of factors $V_{tr}$ and factor loadings $U_{ir}$. The factors are drawn independently from a multivariate normal distribution with means and variances of 1, and covariances of 0.2; the factor loadings are simulated from a first-order autoregressive model with slope coefficient $\phi = 0.3$. I generate ground-truth propensity scores according to the treatment model \eqref{treatment-model}, where the simulated covariate is also generated using a first-order autoregressive model with $\phi = 0.3$.  I focus on the case of a square matrices, $N \times T = 60 \times 60$. For each trial run, I sample $N_{T} = 30$ treated units, and use the ground-truth propensity scores as sampling weights. Each treated unit is assigned an initial treatment period, with the first treatment time set to $a_{i}^{\prime} = 3$. 

\subsection{Empirical data}

\subsubsection{State government finances datasets}

\begin{figure}[htbp]
	\centering
	\begin{subfigure}[b]{0.49\textwidth}
		\centering
		\includegraphics[width=\textwidth]{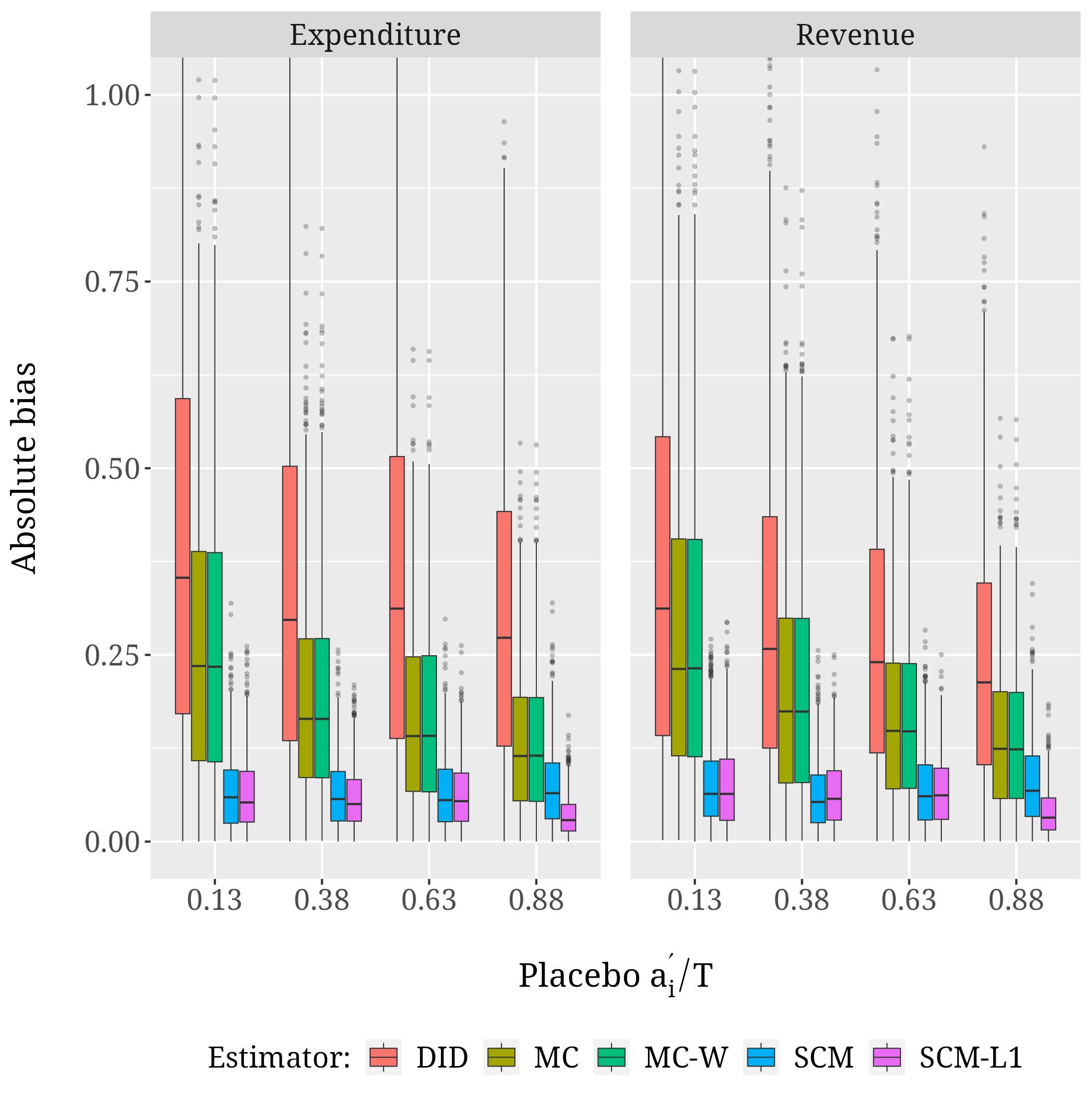}
		\caption{Absolute bias}
		\label{mc_capacity_simulation_placebo_abs_bias}
	\end{subfigure}
	\hfill
	\begin{subfigure}[b]{0.49\textwidth}
		\centering
		\includegraphics[width=\textwidth]{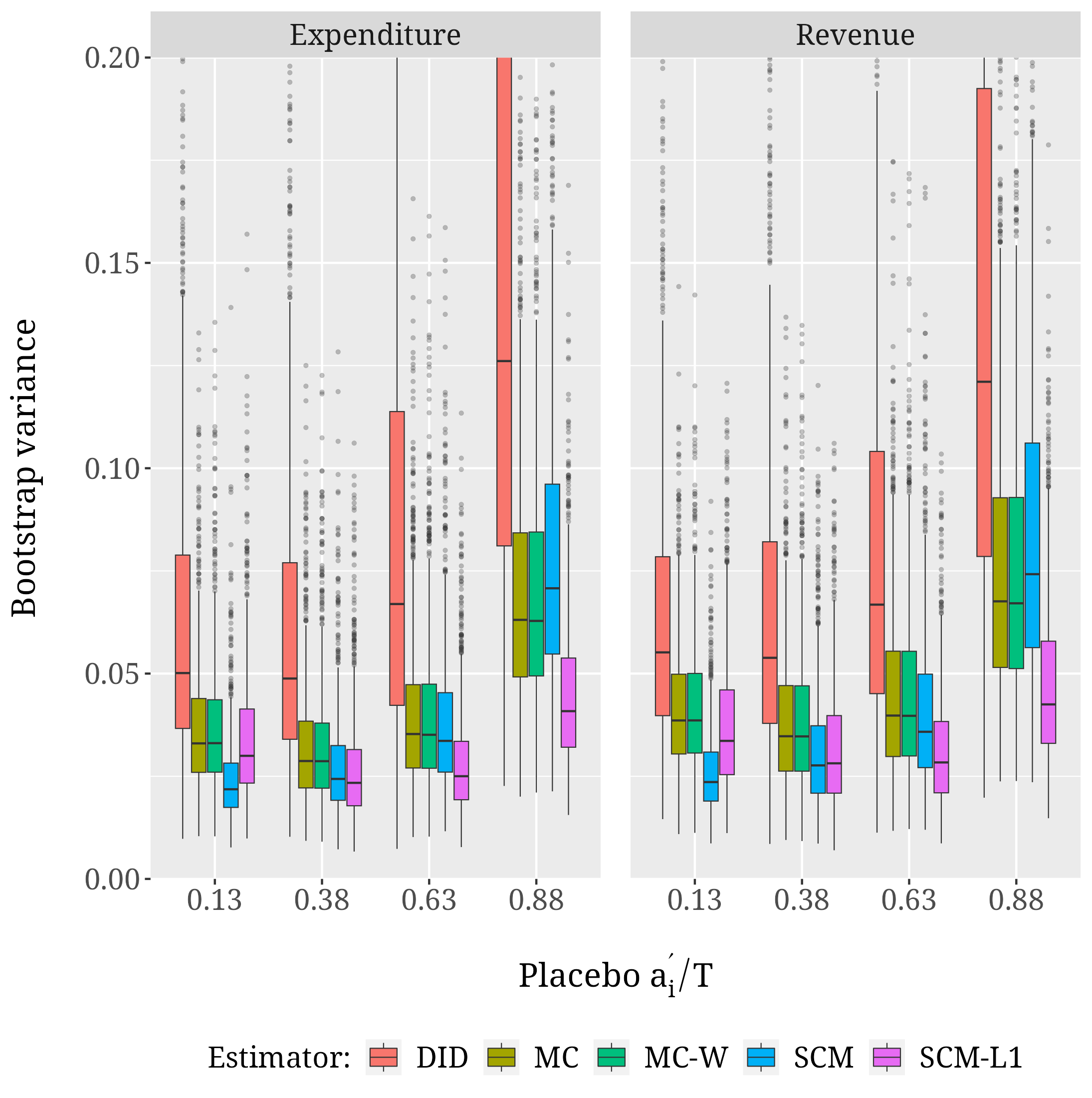}
		\caption{Bootstrap variance}
		\label{simulation-boot-var}
	\end{subfigure}
	
	\caption{Absolute bias and bootstrap variance for state government finances datasets, varying the placebo $a_{i}^{\prime}$.
	\label{mc_capacity_simulation_placebo}}
\end{figure}

I evaluate the performance of the matrix completion estimator on the state government expenditure and revenue datasets described in Section \ref{state-capacity} of the main paper, discarding the treated units and using the control units ($N=18$, $T=203$). Figures \ref{mc_capacity_simulation_placebo_abs_bias} and \ref{simulation-boot-var} present box and whisker plots summarizing the absolute bias and variance, respectively, across 1000 simulation runs on the state government finances datasets. The $x$ axis in each figure is the ratio of the placebo initial treatment time to the number of periods in the placebo data, so higher values represent more training data.

\subsubsection{Synthetic control datasets}

\begin{figure}
	\centering
	\begin{subfigure}[b]{0.49\textwidth}
		\centering
		\includegraphics[width=\textwidth]{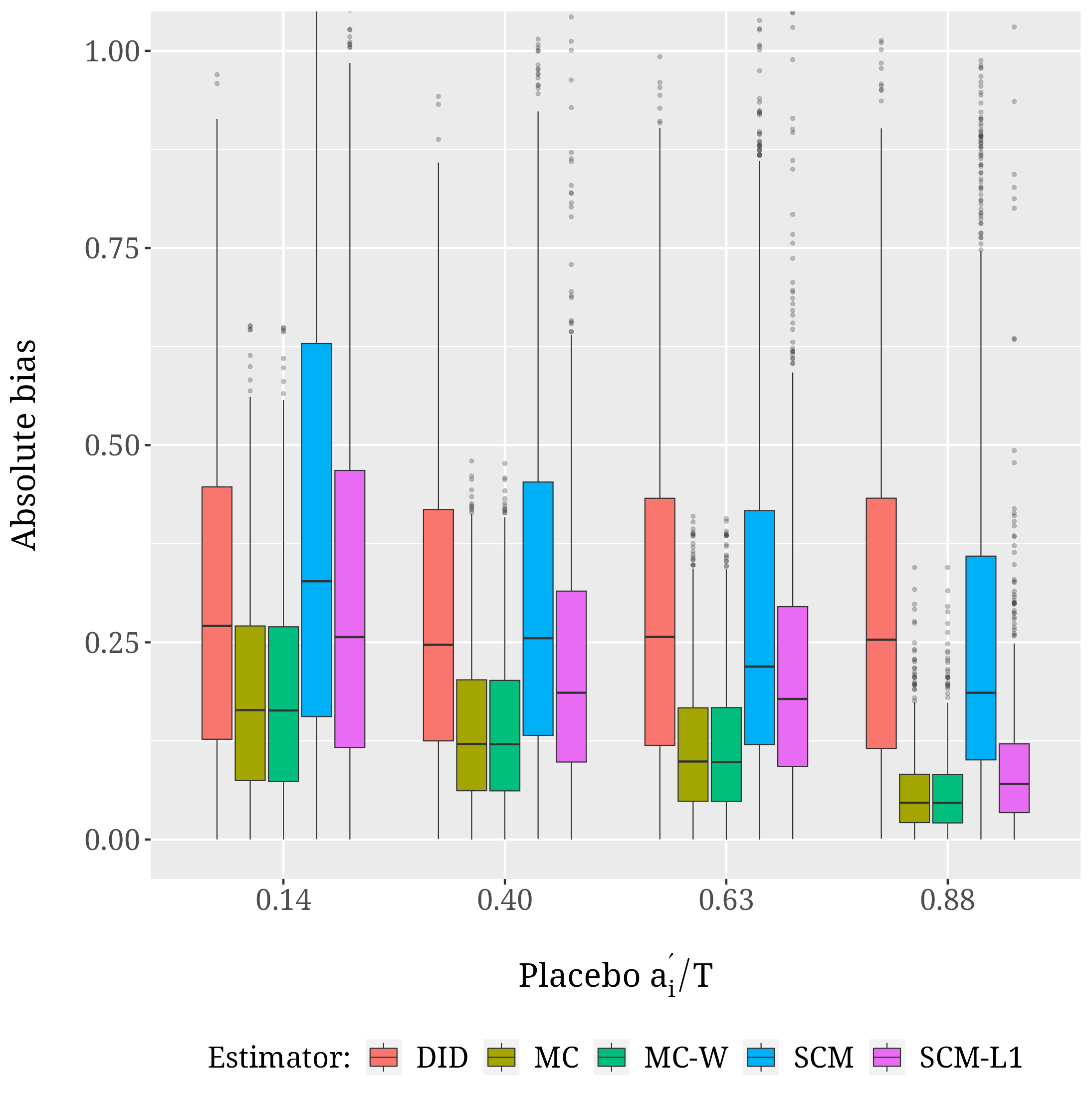}
		\caption{Basque Country terrorism}
		\label{fig:basque_abs_bias}
	\end{subfigure}
\hfill
	\begin{subfigure}[b]{0.49\textwidth}
		\centering
		\includegraphics[width=\textwidth]{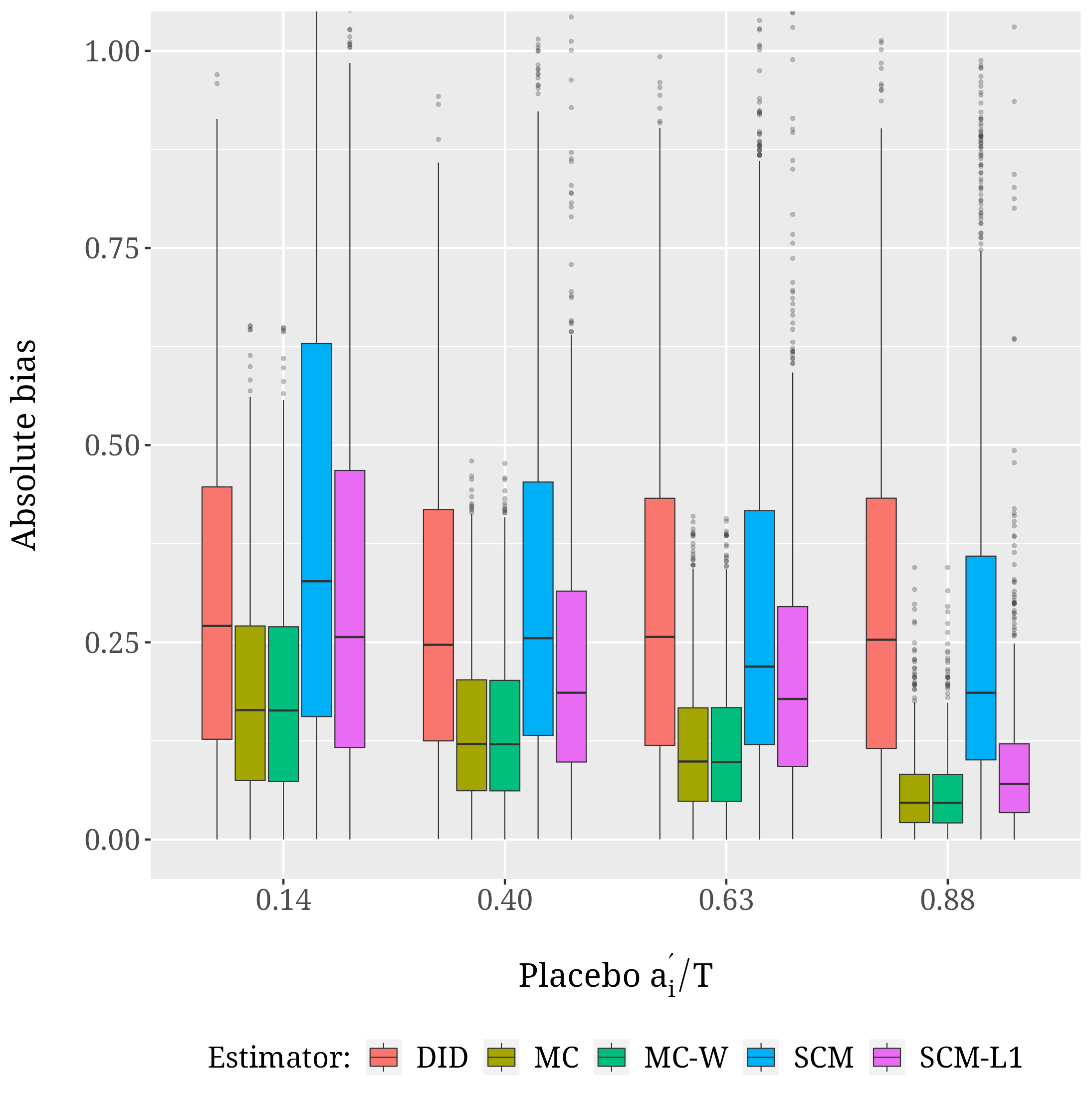}
		\caption{California smoking ban}
		\label{fig:california_abs_bias}
	\end{subfigure}
\vfill
	\begin{subfigure}[b]{0.49\textwidth}
		\centering
		\includegraphics[width=\textwidth]{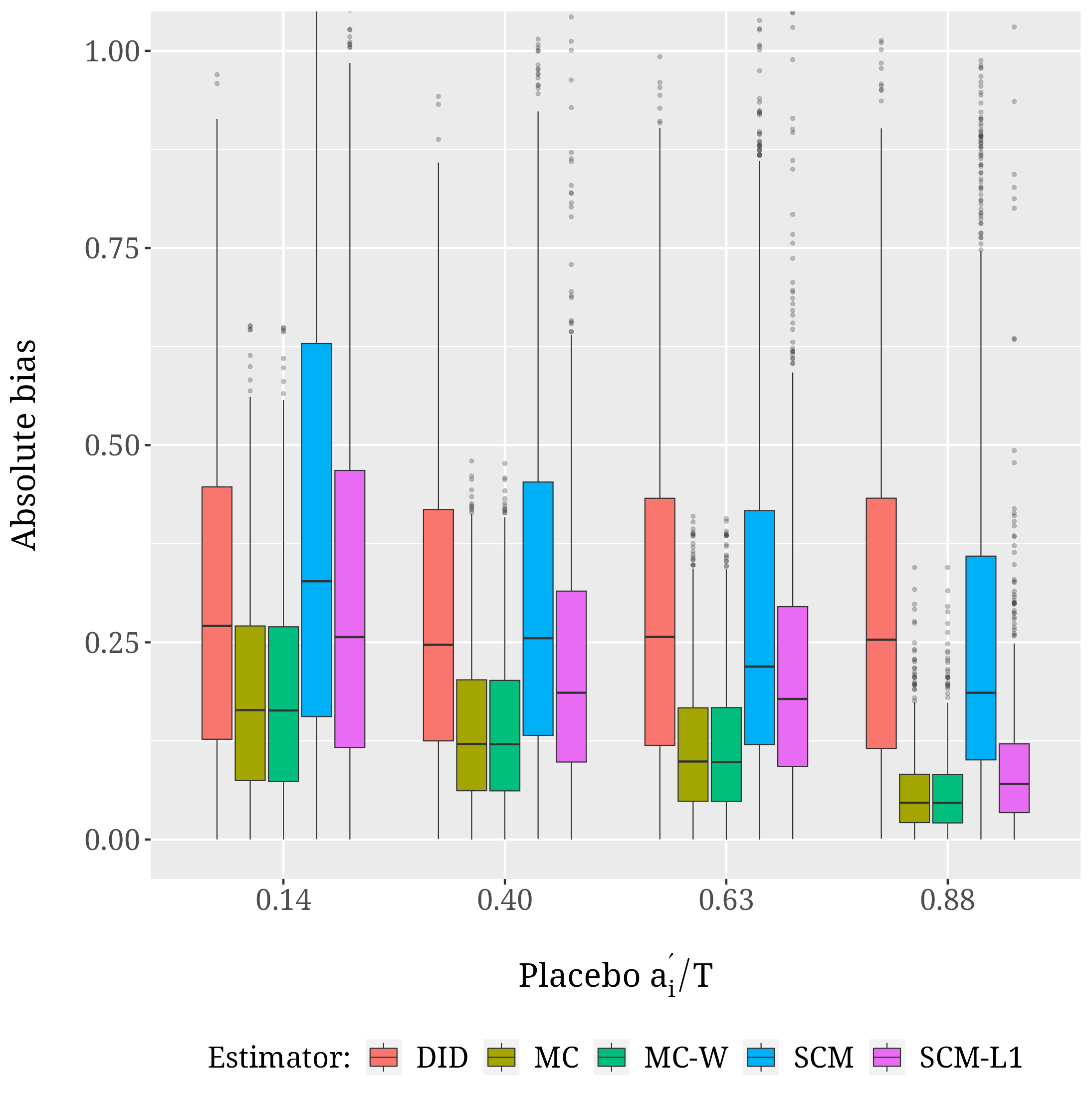}
		\caption{West German reunification}
		\label{fig:germany_abs_bias}
	\end{subfigure}
	\caption{Absolute bias for synthetic control datasets, varying the placebo $a_{i}^{\prime}$.}
	\label{synth_abs_bias}
\end{figure}

Next, I evaluate the performance of the matrix completion estimator on three datasets common to the synthetic control literature, with the actual treated unit removed from each dataset. The three synthetic control datasets originate from \possessivecite{abadie2003economic} study of the economic impact of terrorism in the Basque Country during the late 1960s ($N=16$, $T=43$); \possessivecite{abadie2010synthetic} study of the effects of a large-scale tobacco control program implemented in California in 1988 ($N=38$, $T=31$); and \possessivecite{abadie2015comparative} study of the economic impact of the 1990 German reunification on West Germany ($N=16$, $T=44$). Figure \ref{synth_abs_bias} provides box and whisker plots of the absolute bias for each estimator on the three synthetic control datasets. 

\begin{figure}
	\centering
	\centering
	\begin{subfigure}[b]{0.49\textwidth}
		\centering
		\includegraphics[width=\textwidth]{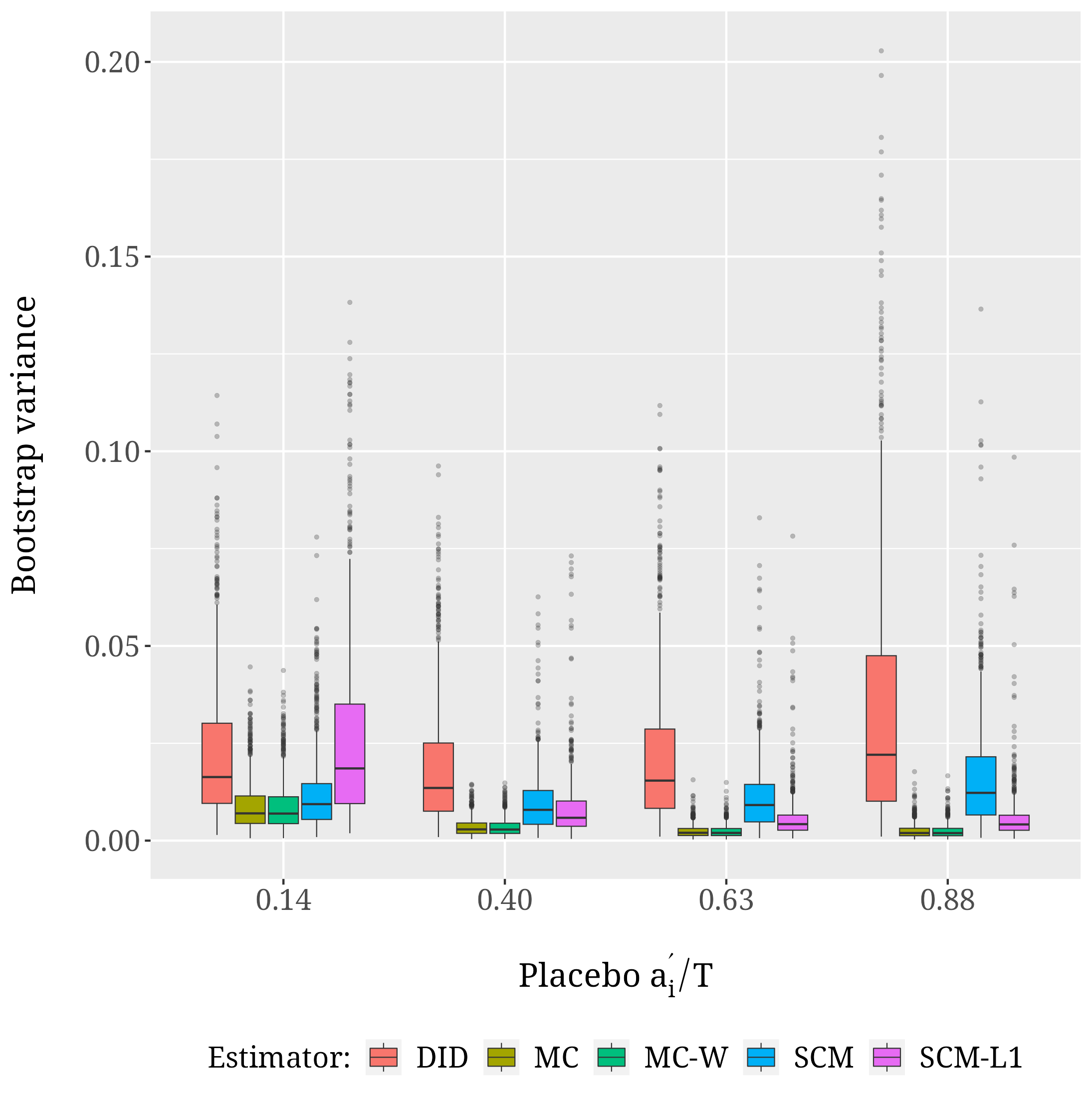}
		\caption{Basque Country terrorism}
		\label{fig:basque_boot_var}
	\end{subfigure}
	\begin{subfigure}[b]{0.49\textwidth}
		\centering
		\includegraphics[width=\textwidth]{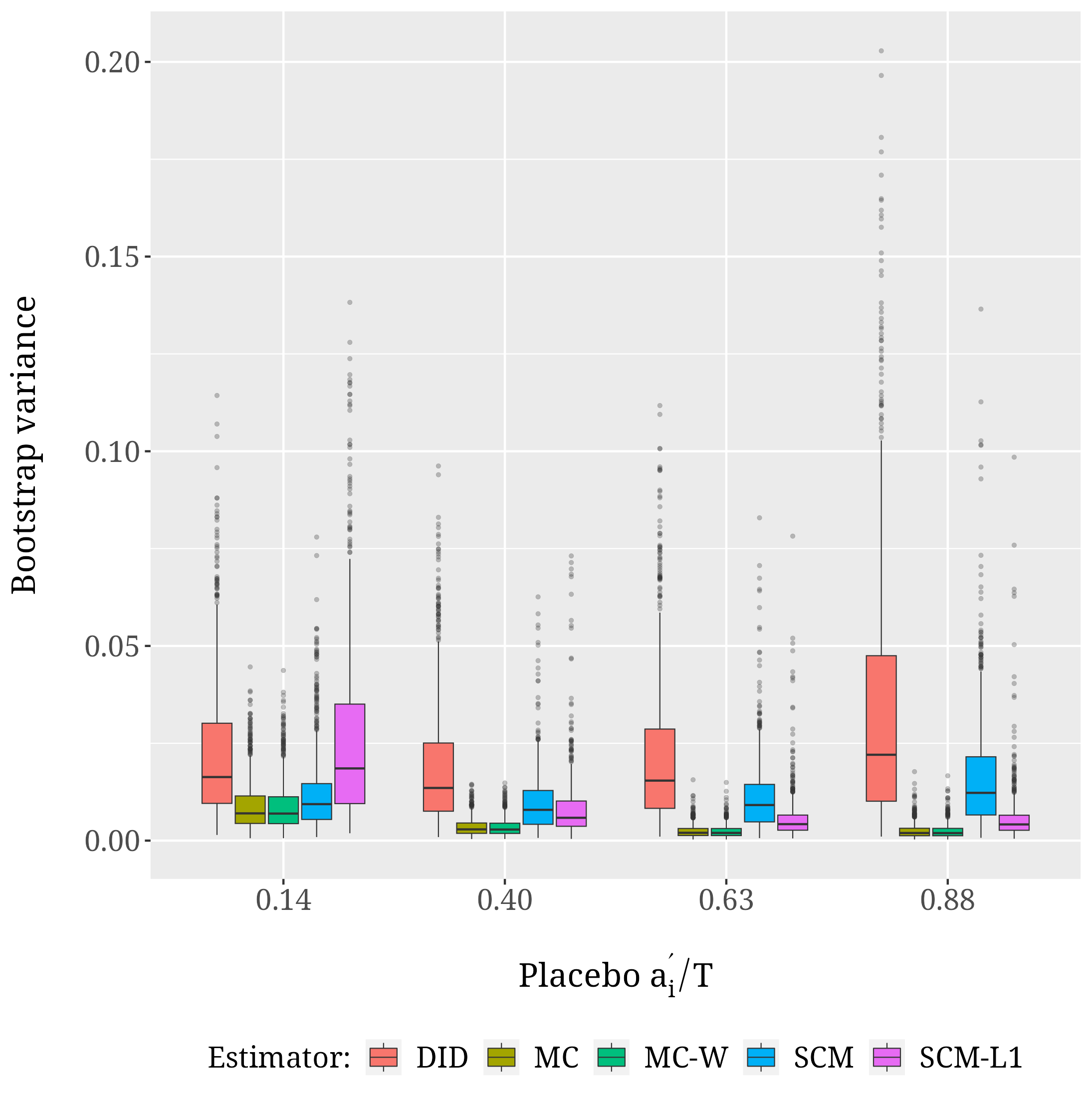}
		\caption{California smoking ban}
		\label{fig:california_boot_var}
	\end{subfigure}
	\begin{subfigure}[b]{0.49\textwidth}
		\centering
		\includegraphics[width=\textwidth]{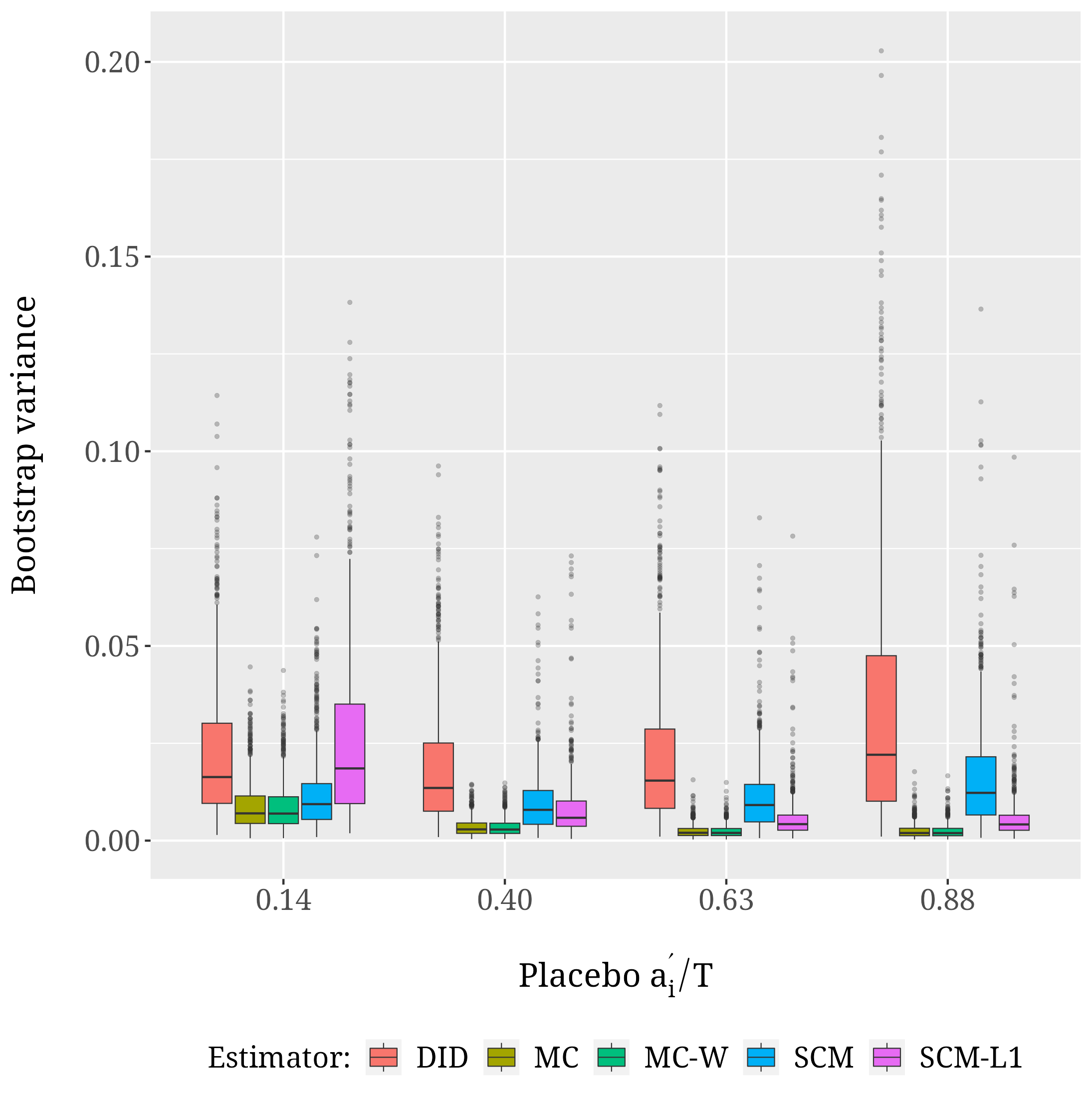}
		\caption{West German reunification}
		\label{fig:germany_boot_var}
	\end{subfigure}
	\caption{Bootstrap variance for synthetic control datasets, varying the placebo $a_{i}^{\prime}$.}
	\label{synth_boot_variance}
\end{figure}

\section{Benchmark estimators} \label{benchmark-estimators}

The following estimators are used for comparison in the no treatment evaluation (Section \ref{placebo-tests}) and main estimates (Section \ref{main-estimates}).

\subsection{Difference-in-differences (DID)}\label{did-binary}

The DID model with binary treatment is specified by \citet{athey2021design}. The outcome under control is modeled as:
\begin{equation}
	Y(\infty)_{it}  = \xi_i + \psi_t + \tau W_{it} + \upsilon_{it},
\end{equation}
where $\{\xi_i\}^N_{i=1}$ are unit-specific fixed effects and $\{\psi_t\}^T_{t=1}$ are time-specific fixed effects. The model is estimated by least squares: 
\begin{equation}
	\argmin _{\tau,\xi_{i},\psi_{t}} \sum_{i=1}^{N} \sum_{t=1}^{T}\left(Y_{i t}-\xi_{i}-\psi_{t}-\tau W_{i t}\right)^{2}.
\end{equation}
I use the implementation of DID in the \texttt{MCPanel} \textsf{R} package \citep{athey2017mcpanel}.

%
%

\subsection{Matrix completion}

Matrix completion methods attempt to impute missing values by solving a convex optimization problem via nuclear norm minimization, even when relatively few values are observed in the data matrix \citep{candes2009exact,candes2010matrix,mazumder2010spectral,recht2011simpler}. I implement two versions of matrix completion estimated by via nuclear norm regularized least squares \citep{athey2021matrix}: with (MC-W) and without (MC) an propensity-weighted loss function. The MC-W outcome model and estimation equation is specified in Eqs. \eqref{eq:mc-Y} and \eqref{eq:mc-opt-prop} of the main paper, respectively. The MC outcome model is the same Eq. \eqref{eq:mc-Y} of the main paper, while the estimation equation does not weight the loss function:

\begin{equation}
	\argmin_{\mathbf{L}, \boldsymbol{\gamma}, \boldsymbol{\delta}} \Bigg[\frac{1}{|\mathcal{O}|} \sum_{(i,t) \in \mathcal{O}} \bigg(Y_{it} - L_{it} - \gamma_{i} - \delta_{t} \bigg)^2
	+ \lambda_L \norm{\mathbf{L}}_\star \Bigg]. \label{eq:mc-opt}
\end{equation}
Both versions are implemented using an extended version of the \texttt{MCPanel} package that includes an option for propensity-weighting the loss function.\footnote{Available at \url{https://github.com/jvpoulos/MCPanel}.}

\subsection{Synthetic control method (SCM)}

The synthetic control method (SCM, \citealp{abadie2010synthetic}) compares a single treated unit with a synthetic control that combines the outcomes of multiple control units on the basis of their pre-intervention similarity with the treated unit. 

\citet{doudchenko2016balancing} and \cite{athey2021matrix} show that the SCM can be interpreted as regressing the pre-treatment outcomes of a single treated unit on the control unit outcomes during the same periods. The parameters estimated on the controls are then used to predict the counterfactual outcomes for a single treated unit, $i=0$:
\begin{align}\label{adh} \nonumber
	\hat{Y}_{0t} &= \sum_{i=1}^{\textrm{N}} \hat{\omega}_i Y_{it}, \qquad \forall \, t= \alpha^{\prime}, \ldots, T,\\
	\textrm{where } \hat{\omega} &=\argmin_{\omega}\sum_{s=1}^{\alpha^{\prime}} \left(Y_{0s}-\sum_{i=1}^{\textrm{N}} \omega_i Y_{0s}\right)^2, \qquad \textrm{s.t. } w_i \geq 0, \sum w_i =1.
\end{align}
A separate model is subsequently fit for each $i, \ldots, N_t$ treated units. Note that Eq. \eqref{adh} imposes the restrictions of the original SCM, namely zero intercept and non-negative regression weights that sum to one. I use the \texttt{MCPanel} implementation with default settings, except I bound gradient values within [-5,5] in order to facilitate convergence.

\subsection{SCM with lasso (SCM-L1)}

The SCM with lasso (SCM-L1, \citealp{doudchenko2016balancing,athey2021matrix}) relaxes the zero-intercept and weight restrictions of Eq. \eqref{adh}. The counterfactual outcomes for treated unit $i=0$ is:
\begin{align}\label{vt} \nonumber
	\hat{Y}_{0t} &= \hat{\mu} + \sum_{i=1}^{\textrm{N}} \hat{\omega}_i Y_{it}, \qquad \forall \, t= \alpha^{\prime}, \ldots, T,\\
	\textrm{where } \left(\hat{\mu}, \hat{\omega}\right) &=\argmin_{\mu, \omega}\sum_{s=1}^{\alpha^{\prime}} \left(Y_{0s}-
	\mu-\sum_{i=1}^{\textrm{N}} \omega_i Y_{0s}\right)^2, 
\end{align}
where a separate model is fit for each $i, \ldots, N_t$ treated units. Intuitively, the generalized SCM is a convex combination of control units with intercept $\mu$ and weight $\omega_i$ for control units $i, \ldots, \textrm{N}$. The model is fit with $N+1$ predictors, including the number of control units and the intercept, and $\alpha^{\prime}, \ldots, T$ observations. 

Eq. \eqref{vt} is estimated by lasso regression \citep{tibshirani1996regression,tibshirani2012strong} in order to reduce the relative size of the predictor set. I use the \texttt{MCPanel} implementation with the strength of the lasso penalty selected among 30 possible values by 5-fold cross-validation.

\clearpage
\section{Empirical application: tables \& figures}\label{tables-figures}

\begin{figure}[htbp]
	\centering
	\includegraphics[width=\textwidth]{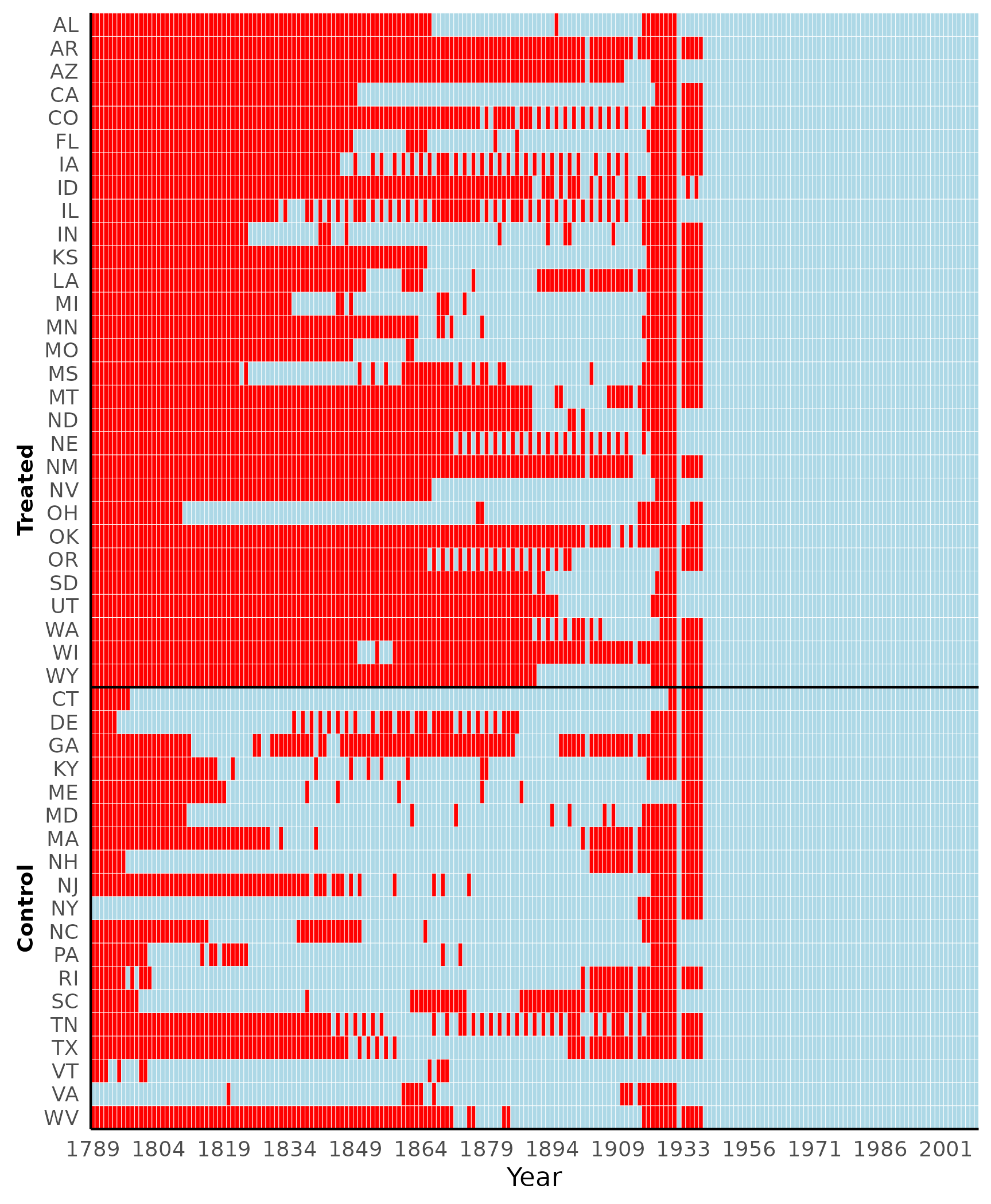}
	\caption{State-year observations in the state government finances datasets that are missing ({\protect\tikz \protect\draw[color={Rred}] (0,0) -- plot[mark=square*, mark options={scale=2}] (0,0);}) or present ({\protect\tikz \protect\draw[color={Rlightblue}] (0,0) -- plot[mark=square*, mark options={scale=2}] (0,0);}). \label{fig:missing-heatmap}}
\end{figure}

\begin{figure}
	\begin{center}
		\includegraphics[width=\textwidth]{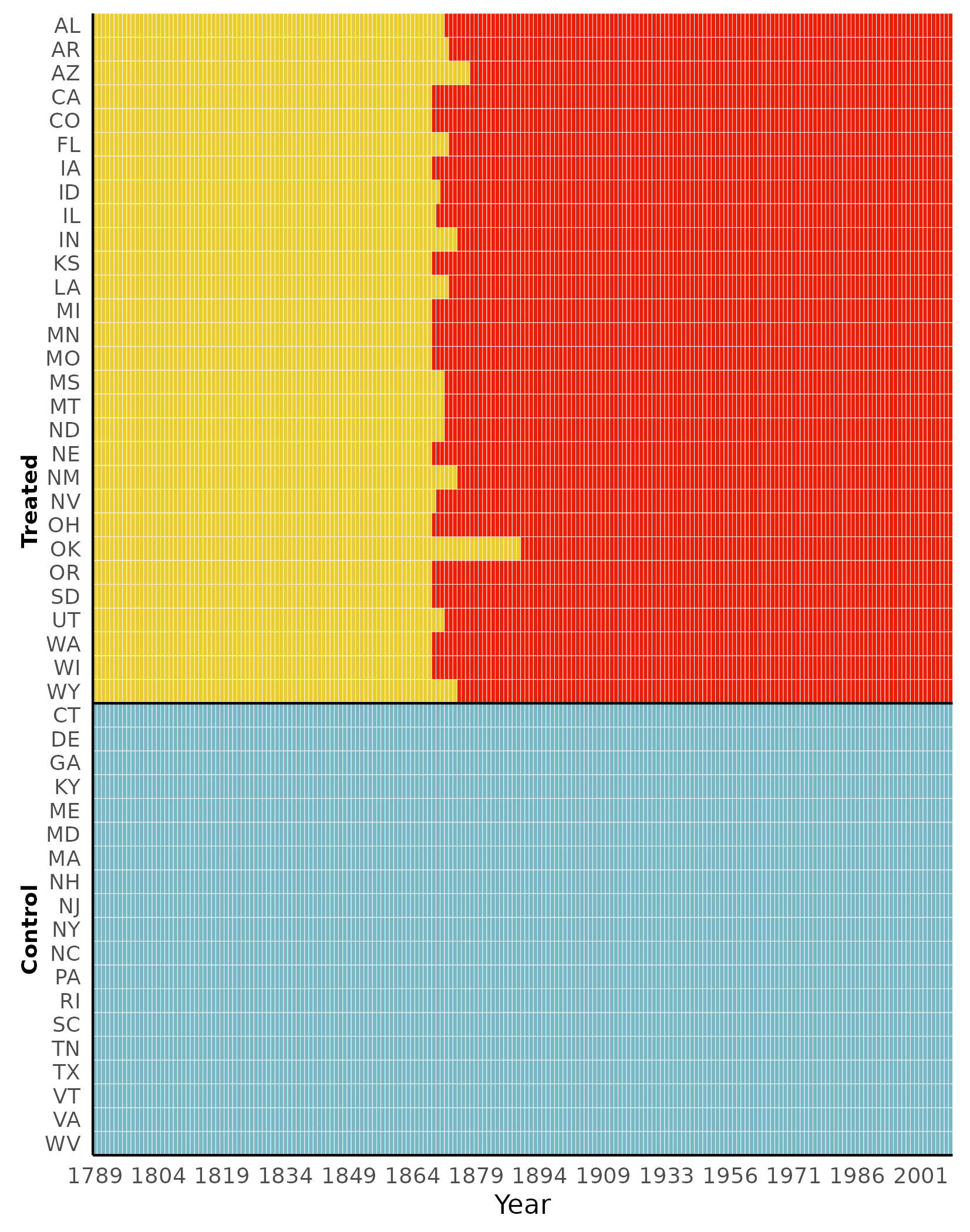}
	\end{center}
	\caption{Treatment status by state and year corresponding to the state government finances datasets.  `Control' are state-land states ({\protect\tikz \protect\draw[color={Darjeeling15}] (0,0) -- plot[mark=square*, mark options={scale=2}] (0,0);})  and `treated' are public land states, before ({\protect\tikz \protect\draw[color={Darjeeling13}] (0,0) -- plot[mark=square*, mark options={scale=2}] (0,0);}) and after ({\protect\tikz \protect\draw[color={Darjeeling11}] (0,0) -- plot[mark=square*, mark options={scale=2}] (0,0);}) treatment. \label{fig:treat-heatmap}}
\end{figure}

\begin{figure}[htbp]
	\centering
	\includegraphics[width=\textwidth]{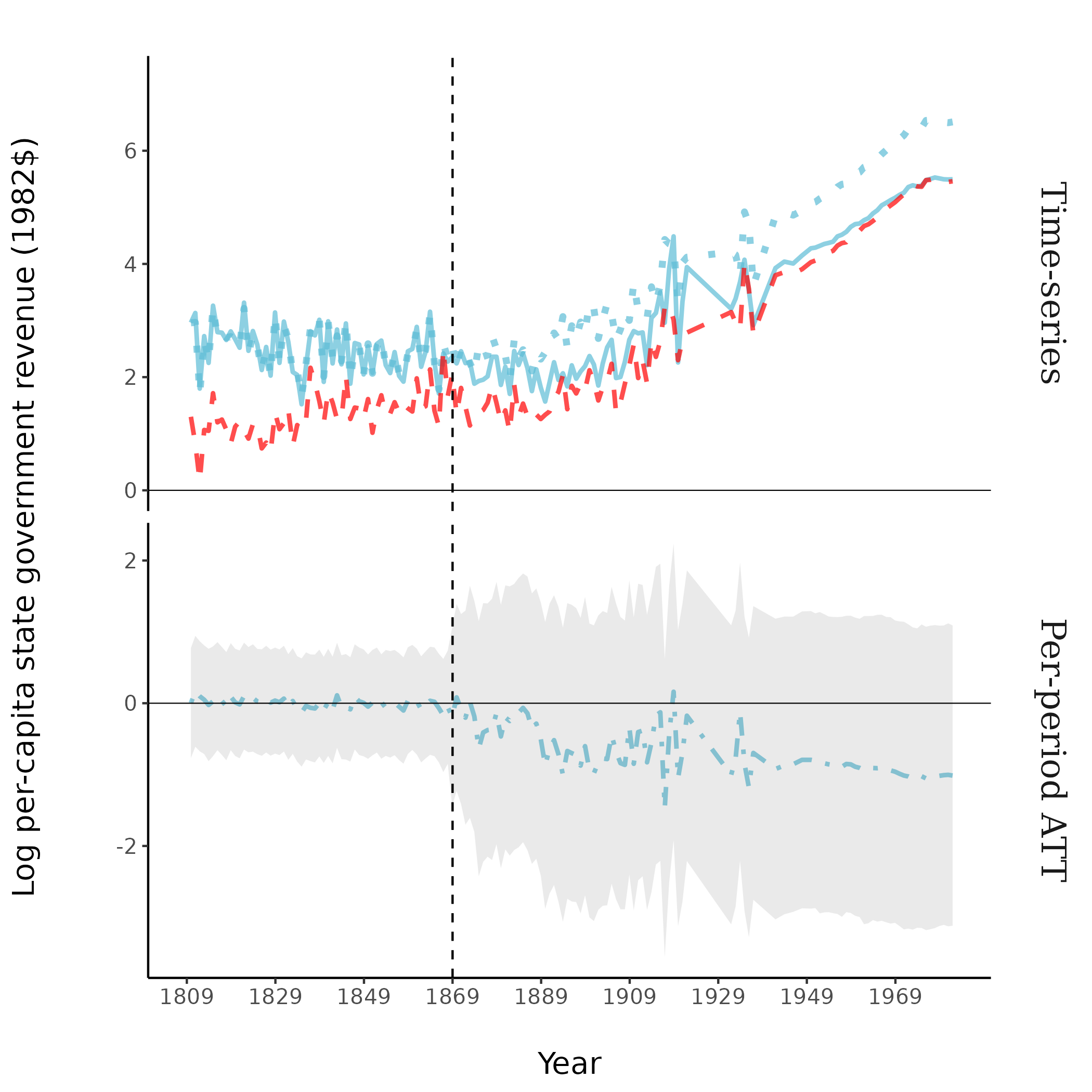}
	\caption{Matrix completion estimates of the effect of the year of initial homestead entry (1869; dashed vertical line) on state government revenue, 1809 to 1982:
		{\color{Darjeeling15}{\sampleline{line width=1mm}}}, factual treated;
		{\color{Darjeeling11}{\sampleline{dashed,line width=0.75mm}}}, factual control;
		{\color{Darjeeling15}{\sampleline{dotted,line width=0.75mm}}}, counterfactual treated;
		{\color{Darjeeling15}{\sampleline{dash pattern=on .7em off .2em on .05em off .2em,line width=0.75mm}}}, $\widehat{\tau}_{t, \infty a_{i}^{\prime}}$.\label{mc-estimates-rev-pc}} 
\end{figure}

\begin{figure*}[t!]
	\centering
	\begin{subfigure}[t]{0.5\textwidth}
		\centering
		\includegraphics[width=\textwidth]{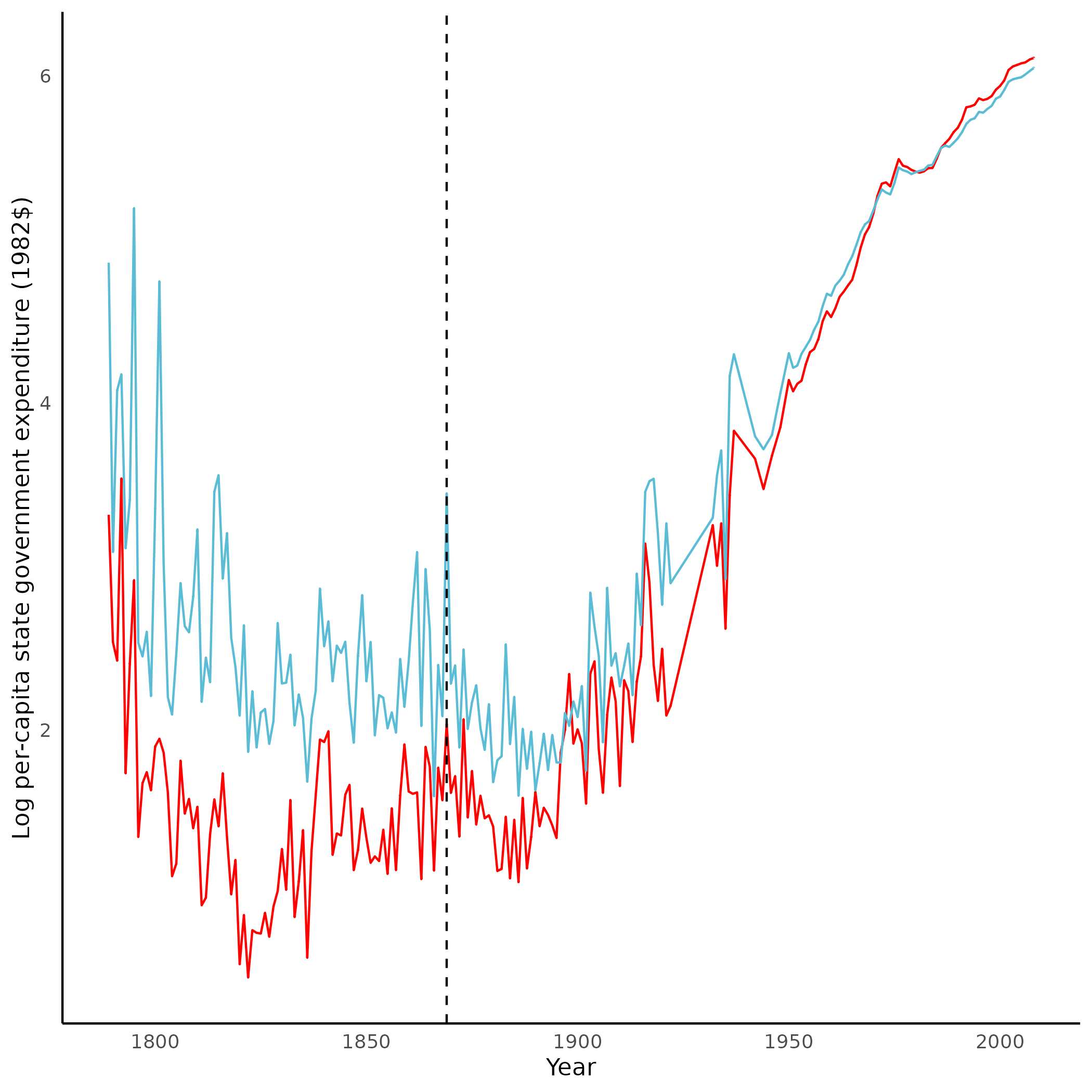}
		\caption{State government expenditure.}
	\end{subfigure}%
	~ 
	\begin{subfigure}[t]{0.5\textwidth}
		\centering
		\includegraphics[width=\textwidth]{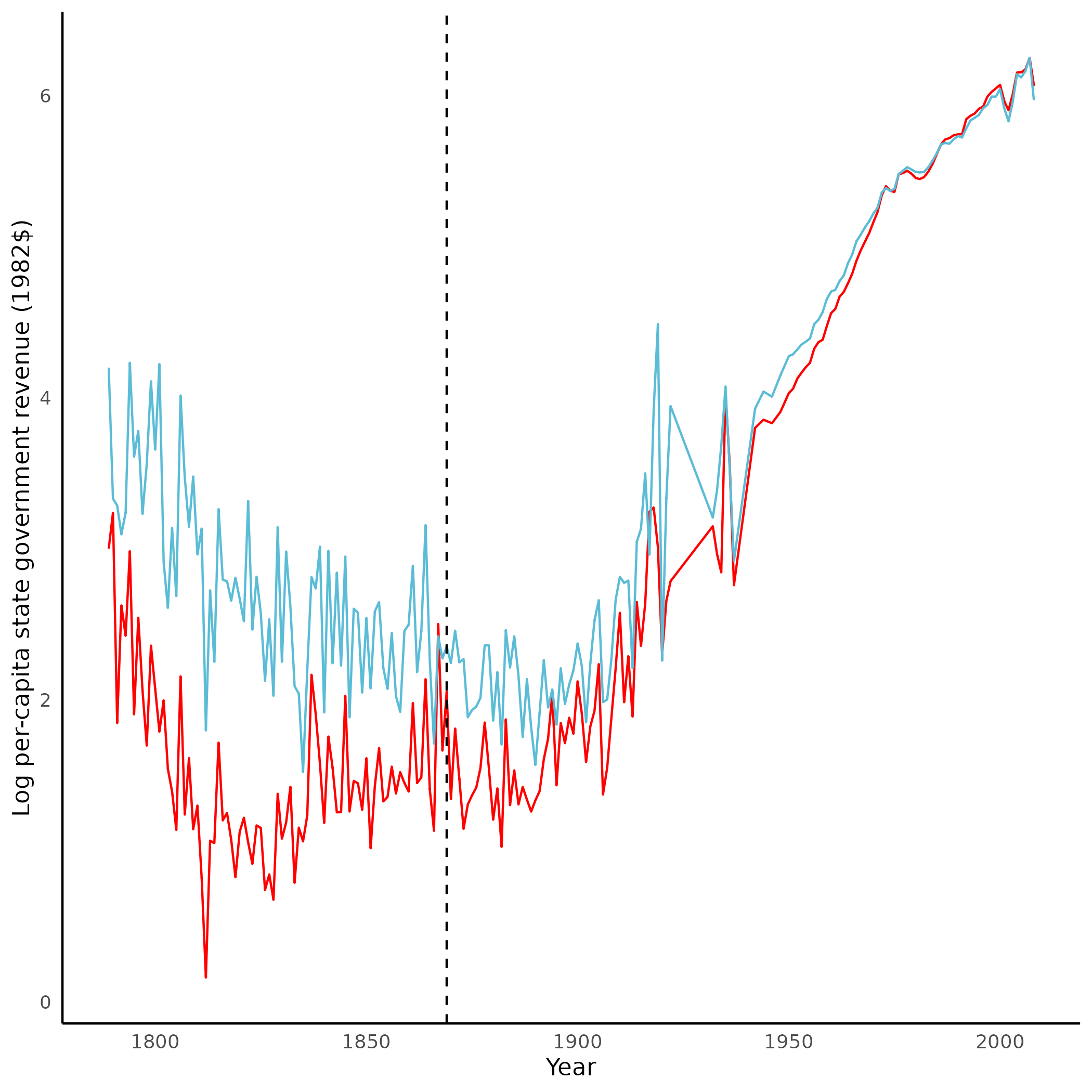}
		\caption{State government revenues.}
	\end{subfigure}
	\caption{Visually assessing the parallel trends assumption for DID estimation with a binary treatment variable: 	{\color{Darjeeling15}{\sampleline{}}}, factual treated;
		{\color{Darjeeling11}{\sampleline{}}}, factual control. The dashed vertical line represents the earliest treatment year, 1869.\label{did-pretest}}
\end{figure*}

\begin{itemize}
	\item Figure \ref{fig:treat-heatmap} shows how each state is categorized in the empirical analysis, and the year of the earliest initial homestead entry for the treated states. 
	\item Figure \ref{fig:missing-heatmap} visualizes the extent of the missing values in the state government finances datasets. 
	\item Figure \ref{fig:ineq-capacity} plots bivariate regression estimates of the relationship between land inequality and state government finances.
	\item Figure \ref{mc-estimates-rev-pc} plots matrix completion estimates of treatment exposure on state government revenue. 
	\item Figure \ref{did-pretest} visualizes the average outcomes of treated and control groups for the purpose of assessing the parallel trends assumption for DID estimation with a binary treatment variable.
	\item  In the no treatment evaluation and in the main analysis, the missing values are imputed by multivariate imputation by chained equations (MICE, \citealp{azur2011multiple}) with predictive mean matching as the imputation method, implemented using the \texttt{mice} \textsf{R} package \citep{buuren2010mice}. Table \ref{mc-sens} reports the results of sensitivity analyses on differently imputed datasets using the following two alternative imputation methods:
		\begin{description}
		\item[MICE-CART] MICE with classification and regression trees (CART) from the \texttt{rpart} package \citep{therneau2022introduction} as the imputation method;
		\item[EM] An expectation--maximization (EM) algorithm based method for imputing missing values in multivariate normal time series, implemented using the \texttt{mtsdi} package with default settings \citep{therneau2022introduction}.
	\end{description}
	In order to avoid train-test set contamination, each imputation method is fit only on the outcome values of the control units. Table \ref{placebo-tests-sens} reports the estimated ATT on differently imputed placebo datasets, and Table \ref{dd-estimates-sens} reports treatment effect estimates using the continuous DID model (Section \ref{DID}) on differently imputed datasets. 
\end{itemize}
 
\begin{table}[htbp]
	\captionsetup{font=normalsize}
	\caption{Estimates of the ATT averaged over the counterfactual period \eqref{eq:ATT-avg} and bootstrap standard errors (in parentheses) on differently imputed datasets.\label{mc-sens}}
	\begin{center}
		\begin{tabular}{@{}lcccccc@{}}
	\toprule 
	& \multicolumn{2}{c}{\emph{All PLS} } & \multicolumn{2}{c}{ \emph{Southern PLS}  } & \multicolumn{2}{c}{ \emph{Western PLS}} \\
	& \multicolumn{1}{c}{ Expenditure } & \multicolumn{1}{c}{ Revenue } & \multicolumn{1}{c}{ Expenditure } & \multicolumn{1}{c}{ Revenue } & \multicolumn{1}{c}{ Expenditure } & \multicolumn{1}{c}{ Revenue } \\
	\multicolumn{3}{l}{ \emph{Imputation method: MICE-CART}  } & \multicolumn{3}{c}{ } & \\
	DID  & -0.66 (0.26) & -0.73 (0.27) & -0.53 (0.37)  & -0.76 (0.41)	&  -0.69 (0.39) &	-0.73 (0.39)	\\
	MC-W & -0.69 (0.21) & -0.63 (0.21) & -0.52 (0.32) & -0.66 (0.33) &  -0.72 (0.33) & -0.62 (0.32)\\
	SCM  & 0.13 (0.14) & 0.08 (0.14)  & -0.11 (0.22) & -0.16 (0.24) & 0.18 (0.25)  & 0.13 (0.18)	\\
	SCM-L1  & 0.16 (0.16) & 0.11 (0.15)	& -0.11 (0.22)  & -0.15 (0.22) &  0.19 (0.25)  & 0.16 (0.24)\\
	\multicolumn{3}{l}{ \emph{Imputation method: EM}  } & \multicolumn{3}{c}{ } & \\
	DID  & -2.29 (0.82)  & -1.48 (0.55)& -2.11 (0.85)  & -1.45 (0.87) &  -2.32 (0.96) &	-1.48 (0.93) \\
	MC-W & -2.04 (0.60) & -1.34  (0.46) & -1.79 (0.71)  & -1.26 (0.74)& -2.09 (0.80)  & -1.36 (0.79)\\
	SCM  & 0.32 (0.38)  & 0.11 (0.31) &  0.12 (0.44) & -0.33 (0.54)	& 0.36 (0.50)  & 0.20 (0.36)	\\
	SCM-L1  & 0.21 (0.42)  & 0.20 (0.33)& -0.01 (0.48)  & -0.21 (0.53) & 0.26 (0.54)   & 0.26 (0.59)\\
	\bottomrule
\end{tabular}
	\end{center}
\end{table}

\begin{table}[htbp]
	\captionsetup{font=normalsize}
	\caption{Placebo ATT estimates and bootstrap standard errors (in parentheses) on differently imputed datasets.\label{placebo-tests-sens}}
	\begin{center}
		\scalebox{.9}{\begin{tabular}{@{}lcccccc@{}} 
	\toprule
		& \multicolumn{3}{c}{ Expenditure } & \multicolumn{3}{c}{ Revenue } \\
	  &  $\Delta=1$ & $\Delta=10$ & $\Delta=25$  &  $\Delta=1$ & $\Delta=10$ & $\Delta=25$  \\
	\multicolumn{3}{l}{ \emph{Imputation method: MICE-CART}  } & \multicolumn{3}{c}{ } & \\
	DID  & -0.35 (0.51) & -0.60 (0.39) & -0.51 (0.28) & -0.04 (0.53)  & -0.61 (0.29) & -0.49 (0.25) \\
	MC-W  & -0.19 (0.43) & -0.52 (0.23) & -0.45 (0.17) & -0.08 (0.38)  & -0.52 (0.19) & -0.27 (0.15)  \\
	SCM  & 0.51 (0.53) & 0.16 (0.24) & 0.79 (0.52) & 0.10 (0.43) & -0.21 (0.24) &  0.09 (0.17)  \\
	SCM-L1  & 0.47 (0.48) & -0.15 (0.22)& 0.12 (0.18) & 0.84 (0.44) & 0.05 (0.20) & 0.22 (0.17)  \\
	\multicolumn{3}{l}{ \emph{Imputation method: EM}  } & \multicolumn{3}{c}{ } & \\
	DID  & -0.81 (0.87) & -1.27 (0.80) & -1.11 (0.63) & -1.75 (1.81)  & -2.11 (1.42) & -1.76 (1.05) \\
	MC-W & -0.41 (0.54) & -0.68 (0.32) & -0.64 (0.30) & -1.71 (1.35)  & -0.59 (0.81) & -1.26 (0.64) \\
	SCM  &  0.89 (0.90) & 0.65 (0.66) & 0.79 (0.52) &  -0.85 (1.47) & -0.16 (1.18) & 0.38 (0.88)  \\
	SCM-L1  & -0.19 (0.55) & -0.16 (0.39)& 0.26 (0.36) & -1.18 (0.88) & -0.04 (0.72) & -0.59 (0.65)  \\
	\bottomrule
\end{tabular}}
	\end{center}
\end{table}

\begin{table}[htp]
	\caption{Continuous DID estimates of the effect of (log) per-capita homestead patents on differently imputed state government finance datasets. \label{dd-estimates-sens}}
	\begin{center}
		\resizebox{.9\width}{!}{	\begin{tabular}{@{}lcccc@{}}
		\toprule
		\emph{Imputation method:} & \multicolumn{2}{c}{MICE-CART} & \multicolumn{2}{c}{EM} \\
		\emph{Outcome:}								& Expenditure             & Revenue   & Expenditure             & Revenue               \\ \hline \\
		Treatment effect ($\hat{\phi}$) &  -0.04 				  &  -0.04   	&  -0.04 				  &  -0.06 				       \\
										&  (0.002) 	  	  & (0.002)   	&  (0.002) 	  	  & (0.003)  		         \\
									    &                   	  &             &                   	  &          		         \\
		Adjusted $\text{R}^2$           & 0.568          		  & 0.577         & 0.452          		  & 0.320                    \\
		$\text{N}$                      & 8,618                   & 8,618         & 8,618                   & 8,618                   \\
		 \bottomrule
	\end{tabular}}
	\end{center}
\end{table}

\clearpage
\bibliographystyle{asa}
\begin{singlespace}
	\bibliography{references}
\end{singlespace}

\itemize